\documentclass[12pt,letterpaper]{article}

\usepackage{amsmath}
\usepackage{amsfonts}
\usepackage{amssymb}
\usepackage{graphicx}
\newcommand{\PRD}[3]{\emph{ Phys.~Rev.} \textbf{D#1} (#2) #3}

\begin{document}


\begin{flushright} {\footnotesize MIT-CTP-3623}  \end{flushright}
\vspace{5mm} \vspace{0.5cm}
\begin{center}

\def\thefootnote{\fnsymbol{footnote}}

{\Large \bf Hierarchy from Baryogenesis} \\[1cm]
{\large Leonardo Senatore\footnote{This work is supported in part
by funds provided by the U.S. Department of Energy (D.O.E) under
cooperative research agreement DF-FC02-94ER40818}}
\\[0.5cm]

{\small
\textit{Center for Theoretical Physics, \\
Massachusetts Institute of Technology, Cambridge, MA 02139, USA}}
\end{center}
\vspace{.2cm}

 \vspace{.8cm}

\hrule \vspace{0.3cm}
{\small  \noindent \textbf{Abstract} \\[0.3cm]
We study a recently proposed mechanism to solve the hierarchy
problem in the context of the landscape, where the solution of the
hierarchy problem is connected to the requirement of having
baryons in our universe via Electroweak Baryogenesis. The phase
transition is triggered by the fermion condensation of a new gauge
sector which becomes strong at a scale $\Lambda$ determined by
dimensional transmutation, and it is mediated to the standard model
by a new singlet field. In a ``friendly'' neighborhood of
 the landscape, where only
the relevant operators are ``scanned'' among the vacua,
baryogenesis is effective only if the higgs mass $m_h$ is
comparable to this low scale $\Lambda$, forcing $m_h\sim\Lambda$,
and solving the hierarchy problem. A new CP violating phase is
needed coupling the new singlet and the higgs field to new matter
fields. We study the constraints on this model given by
baryogenesis and by the electron electric dipole moment (EDM), and we briefly
comment on gauge coupling unification and on dark matter relic
abundance. We find that next generation experiments on the EDM
 will be sensitive to essentially the entire viable region
of the parameter space, so that absence of a signal would effectively rule out
the model.

\noindent

\vspace{0.5cm}  \hrule

\def\thefootnote{\arabic{footnote}}
\setcounter{footnote}{0}


\section{Introduction}

The most problematic characteristic of the Standard Model (SM)
seems to be the smallness of its superrenormalizable couplings,
the higgs mass $m^2_h$, and the cosmological constant $\Lambda_c$,
with respect to the apparent cutoff of the theory, the Planck mass,
$M_{pl}$. These give rise respectively to the hierarchy,
and cosmological constant problems. According to the naturalness
hypothesis, the smallness of these operators should be understood
in terms of some dynamical mechanism, and this has been a
driving motivation in the last two decades in the high
energy theory community.

In the last few years, several things have changed.

Concerning the hierarchy problem, experimental investigation has
shown that already many of the theoretically most attractive
possibilities for the stability of the weak scale, such as
supersymmetry, begins to be disfavored, or present at least some
fine tuning issues \cite{Barbieri:1999tm,Giusti:1998gz}.

Concerning the cosmological constant problem, there was the hope
that some symmetry connected to quantum gravity would have forced
the cosmological constant to be zero. However, first, cosmological observations
 have shown evidence for a non zero cosmological
constant in our present universe
\cite{Spergel:2003cb,Perlmutter:1998np}; second, from the string
theory point of view, two main things have occurred: on the one
hand, consistent solutions with a non zero cosmological constant
have been found
\cite{Bousso:2000xa,Giddings:2001yu,Curio:2001qi,Maloney:2002rr,Kachru:2003aw},
and, on the other hand, it has been becoming more and more clear
that string theory appear to have
 a huge landscape of vacua, each one with
different low energy parameters
\cite{Susskind:2003kw,Douglas:2003um}.

If the landscape is revealed to be {\it real}, it would force a
big change in the way physics has to be done, and some deep
questions may find a complete new answer. In particular, it is
conceivable that some characteristics of our low energy theory are
just accidental consequences of the vacua in which we happen to
be. This is a very big step from the kind of physics we are used
to, and its consequences have been explored recently in
\cite{Arkani-Hamed:2005yv}. There , it is shown that the presence
of a landscape offers a new set of tools to address old questions
regarding the low energy effective theory of our universe. On one
hand, there are statistical arguments, according to which we might
explore the statistically favored characteristics for a vacuum in
the landscape
\cite{Douglas:2003um,Dvali:2003br,Ashok:2003gk,Denef:2004ze,Giryavets:2004zr,Conlon:2004ds,
DeWolfe:2004ns,Kumar:2004pv,Dvali:2004tm,Blumenhagen:2004xx,Denef:2004cf}.
On the other hand, there are also some selection rules due to
anthropic reasoning, which we might have to impose on the vacuum,
in order for an observer to be present in it. Pioneering in this,
is Weinberg's ``structure principle'' \cite{Weinberg:1987dv},
which predicts the right order of magnitude for the cosmological
constant starting from the requirement that structures should have
had time to form in our universe in order for life to be present
in it.

At this point, it is necessary to speak about predictivity for
theories formulated with a lot of different vacua, as it occurs in
the landscape. There are two important points that greatly affect the
predictivity of the theory. First, we must know how the
characteristic of the low energy theory change as we scan among
the vacua. Second, we must know also how the probability of
populate such a vacuum changes among the vacua, also considering
the influence of the cosmological evolution. These are very deep
questions whose answer in general requires a full knowledge of the
UV theory, and we do not address them here. However, there is
still something we can do. In fact, as it was pointed out in
\cite{Arkani-Hamed:2005yv}, theories with a large number of vacua
can be described in an effective field theory approach. In such a
study, it was shown that it is very natural for parameters not to
effectively scan in the landscape, unless their average value in
the landscape is null. So, it is reasonable to assume that some
parameters in the landscape do scan, and some do not. Among the
parameters of the theory, the relevant operators have a particular
strong impact on the infrared properties of the theory. But
exactly because of this, if they are the only ones to scan, their
value can be fixed by environmental arguments. This is however
true only if the marginal couplings do not scan. For this reason,
a neighborhood of the landscape in which only the relevant
parameters are scanning, is called a ``friendly neighborhood'',
because it allows to fix the value of the relevant couplings with
environmental reasoning, while the marginal parameters do not
scan, and so they are analyzed with the usual instruments of
physics, judging them on the basis of their simplicity and of
their correlations. We do not know if the true string theory landscape have
these properties. However, there is some
phenomenological motivation to expect that this could well be. In
fact, in the physics connected to the standard model, we have been
able to successfully address question concerning the marginal
parameters with dynamical reasoning, deducing some striking
features as gauge coupling unification, chiral symmetry breaking,
or the weakly interacting dark matter, while, on the contrary, we
have been having big troubles concerning the problems connected to
the relevant coupling in the standard model, the cosmological
constant, and the higgs scale, both of which arise large fine
tuning issues. In this paper, we want to address the hierarchy
problem of the electroweak scale assuming we are living in such a
"friendly neighborhood" of the landscape. Because of this, we
concentrate a little more on the predictivity of a
phenomenological model based on this. As it is clear, predictivity
in the landscape is very much enhanced if there is an infrared
fragile feature \cite{Arkani-Hamed:2005yv} which must be realized
in the universe in order for it not to be lethal, and which is
however difficult to realize in the scanned landscape. Exactly the
fact that a necessary fragile feature is difficult to realize
gives a lot of constraints on the possible vacua in which we
should be, or, in other words, on the parameters of the theory,
making then the model predictive. An example of a fragile feature
is the presence of structures in the universe. According to
Weinberg's argument \cite{Weinberg:1987dv}, if we require the fact
that in the universe there should be structures, than the value of
the cosmological constant is very much tighten to a value close to
the observed one. The presence of structures can then also be used to anthropically require 
the lightness of the higgs field in the case dark matter particles
receive mass from electroweak symmetry breaking \cite{Arkani-Hamed:2005yv,Schuster:2005ck}.

In this paper, we concentrate on another fragile feature which we
consider necessary for the development of any sort of possible
life in the universe, and which is a necessary completion on
Weinberg's structure principle. In the absence of baryons, the
dark matter would just form virilized structures, and not clumped
structures, which instead are necessary for the development of
life. We can
 construct a model where baryons are explicitly a
fragile feature. Then, we can connect the solutions of the problem
of baryons to the hierarchy problem with a simple mechanism which
imposes a low energy electroweak scale in order for baryons to be
present in the universe. This is the line of thought implemented
in \cite{Arkani-Hamed:2005yv}, which is the model on which we
focus in this paper. As we will explain in detail in the next
section, the mechanism is based on the fact that baryogenesis is
possible in the model only if the electroweak scale is close to a
hierarchical small scale. The hierarchical small scale is
naturally introduced setting it equal to the scale at which a new
gauge sector of the theory becomes strong through dimensional
transmutation. This mechanism naturally provides a small scale. In
order to successfully implement baryogenesis, we will also need to
add some more CP violation. This will force the presence of new
particles and new couplings. The model becomes also predictive:
the lightest of these particle will naturally be a weakly
interacting particle at the weak scale, and so it will be a very
good candidate for dark matter; further, the new CP violating
terms will lead to the prediction of a strong electron electric
dipole.

The main purpose of this paper is to investigate in detail the
mechanism in which baryogenesis occurs in this model, and the
consequent predictions of the model. In particular, we will find
that next generation experiments of the electron Electric Dipole
Moment (EDM), together with the turning on of LHC,  are going to explore the entire
viable region of the parameter space, constraining it to such a
peculiar region of the parameter space, so that absence of a
signal would result in ruling out the model.

The paper is organized as follows: in sec. II, we explain in
detail the model; in sec. III we study the amount of baryons
produced; in sec. IV we determine the electron EDM; in sec. V we
briefly comment on Dark Matter and Gauge Coupling Unification ;
and in sec. VI, we draw our conclusions.

\section{Hierarchy from Baryogenesis}
In this section, following \cite{Arkani-Hamed:2005yv}, we show how
we can connect the electroweak scale to a scale exponentially
smaller than the cutoff.

As the presence of baryons is a necessary condition for the
formation of clumped structures, it is naturally to require that
the vacuum in which we are should allow for the formation of a net
baryon number in the universe. This is not a trivial requirement.
In fact, in the early hot phase of the universe, before the
electroweak phase transition, baryon number violating interactions
through weak sphalerons were unsuppressed, with a rate given
approximately by \cite{ew_sphal}:
\begin{equation}
\Gamma_{ws}=6k\alpha^5_w T\label{unbroken_sphaleron_rate}
\end{equation}
where $k\simeq 20$, with the consequence of erasing all
precedently generated baryon asymmetry (if no other macroscopic
charges are present). However, at the electroweak phase
transition, all the Sakharov's necessary conditions
\cite{Sakharov:1967dj} for generating a baryon asymmetry are
satisfied, and so it is possible in principle to generate a net
baryon number at the electroweak phase transition, through a
process known in the literature as electroweak baryogenesis (see
\cite{Riotto:1998bt,Riotto:1999yt} for two nice reviews). However,
it is known that the SM electroweak phase transition can not, if
unmodified, generate the right amount of baryons for two separate
reasons. On one hand, CP violating interactions are insufficient
\cite{Gavela:1994dt}, and, on the other hand, the phase transition
is not enough first order for preventing weak sphalerons to be
active also after the phase transition \cite{Rummukainen:1998as}.
Since this last point is very important for our discussion, let us
review it in detail.

Let us suppose in some early phase of the universe we have
produced some initial baryon number $B$ and lepton number $L$, and no other macroscopic
charge. In particular $B-L=0$. The quantum number $B+L$ is anomalous, and the equation for
the abundance of particles carrying $B+L$ charge is given by:
\begin{equation}
\frac{d\left(\frac{n_{B+L}}{s}\right)}{dt}=-\Gamma\left(\frac{n_{B+L}}{s}
\right)\label{destr}
\end{equation}
where $n_{B+L}$ is the number of baryons and leptons per unit volume, $s$ is the
entropy density $\sim T^3$, and $\Gamma$ is related to the
sphaleron rate \cite{Riotto:1998bt}, $\Gamma\sim\frac{13}{2}N_f T
e^{-(\frac{4\pi}{g_2}) (\frac{v(T)}{\sqrt{2}T})}$, where
$\frac{v(T)}{\sqrt{2}}$ is the temperature dependent higgs vev
($v(T_0)=246$ GeV as measured in our vacuum), $N_f$ is the number
of fermionic families, and $g_2$ is the SU(2) weak coupling. The
reason why reactions destroying baryons are faster than reactions
creating them is due to the fact that the relative reaction rate
goes as:
\begin{equation}
\frac{\Gamma_+}{\Gamma_-}\sim e^{-\Delta f}
\end{equation}
where $\Delta f$ is the difference in free energy, and $\Gamma_\pm$ are the
sphaleron rates in the two directions. Now, it is easy to see that the free
energy grows with the chemical potential $\mu_B$, which then grows with the
number of baryons $n_B$. In the limit of small difference, we then get
eq.(\ref{destr}).

This differential equation can be integrated to give:
\begin{equation}
\left(\frac{n_{B+L}}{s}\right)_{{\rm final}}\simeq\left(\frac{n_{B+L}}{s}\right)_{{\rm
initial}}
{\rm Exp}\left(-\frac{ N_f}{g_*}\frac{\sqrt{2}m_{pl}}{v(T_c)}
e^{-\left(\frac{4\pi}{g_2}\frac{v(T_c)}{\sqrt{2}T_c}\right)}\right)\label{wash_out}
\end{equation}
where $T_c$ is the critical density, and where $g_*$ is the number of
effective degrees of freedom ($\sim 55$).

If in the universe there is no macroscopic lepton number, than
clearly at present time we would have no baryons left, unless:
\begin{equation}
\frac{N_f}{g_*}\frac{\sqrt{2}m_{pl}}{v(T_c)}e^{-\left(\frac{4\pi}{g_2}
\frac{v(T_c)}{\sqrt{2}T_c}
\right)}\lesssim1 \label{suppression}
\end{equation}
which roughly implies the constraint:
\begin{equation}
\frac{v(T_c)}{T_c}\gtrsim1\label{wash_out_const}
\end{equation}

This is the so called "baryons wash out", and the origin of the
requirement that the electroweak phase transition should be
strongly first order.

Note that this is the same condition we would get if we required
sphaleron interactions not to be in thermal equilibrium at the end
of the phase transition:
\begin{equation}
\frac{\Gamma}{T^3_c}<H
\end{equation}
implies roughly:
\begin{equation}
T_c \ e^{-\frac{4\pi}{g_2}\frac{v(T_c)}{\sqrt{2}T_c}}<\frac{T^2_c}{M_{pl}}
\end{equation}
and so
\begin{equation}
\frac{v(T_c)}{T_c}\gtrsim 1
\end{equation}
Also note that the requirement for the sphalerons not to be in
thermal equilibrium already just after the phase transition is
necessary, as otherwise we would have a baryon symmetric universe,
which leads to a far too small residual relic density of baryons.

Later on, when we shall study electroweak baryogenesis, we shall
get a number for the baryon number. In order to get then the
baryon number at, let us say, Big Bang Nucleosynthesis (BBN), we
need to multiply that number by the factor in eq.(\ref{wash_out}).
In the next sections, we shall consider the requirement in
eq.(\ref{wash_out_const}) fulfilled, and we will consider
completely negligible the wash out from the sphalerons coming from
eq.(\ref{wash_out}).

Now, let us go back to the higgs phase transition, and let us
assume that in the neighborhood of the landscape in which we are,
all the high energy mechanisms for producing baryons have been
shut down. This is easy to imagine if, for example, the reheating
temperature is smaller than the GUT scale. We are then left with
the only mechanism of electroweak baryogenesis. From the former
discussion, it appears clear that there should be some physics
beyond the SM to help to make the phase transition strong enough.

We may achieve this by coupling the higgs field to a singlet $S$, with the
potential equal to:
\begin{equation}
V=\lambda S^4+\lambda_h(h^{\dag} h-\tilde{\lambda} S^2)^2+\frac{1}{2}
m^2_S S^2+m^2_h
h^{\dag} h
\end{equation}
where we have assumed a symmetry $S\rightarrow -S$. We can couple this field
to two fermions $\Psi,\Psi^c$, which are charged under a non-Abelian gauge
group through the interaction
\begin{equation}
k_S S\Psi\Psi^c
\end{equation}
In order to preserve the symmetry $S\rightarrow-S$, we give the
fields $\Psi,\Psi^c$ charge $i$. We can then assume that this
sector undergos confinement and chiral symmetry breaking at its
QCD scale determined by dimensional transmutation
\begin{equation}
<\Psi\Psi^c>\sim \Lambda^3
\end{equation}
which is naturally exponentially smaller than the cutoff of the theory.
We assume that this phase transition is first order, so that
departure from thermal equilibrium is guaranteed.

Now, following our discussion in the introduction, suppose that,
scanning in the landscape, the only parameters which are
effectively scanned are the relevant couplings $m_h$ and $m_S$. If
then we must have baryons in our universe so that clumped
structures can form, than we need to be in the vacuum in which
these two parameters allow for a strong enough first order phase
transition in the electroweak sector. So, we have to require that
this phase transition triggers the electroweak
 phase transition. It is clear
 that this can only be if it triggers
a phase transition in the $S$ field, which is possible then only
if $m_S$ is of the order of $\Lambda$. Finally, the phase
transition in $S$ can trigger a  strong first order phase
transition in the higgs field only if again $m_h\sim m_S\sim
\Lambda$ (for a more detailed discussion, see next subsection).
So, summarizing, we see that, the requirement of having baryons in
the universe forces the higgs mass to be exponentially smaller
than the cutoff, solving in this way the hierarchy
problem.

In order to produce baryons, we still need to improve the CP
violating interactions, that in the SM are not strong enough. We
can minimally extend the introduced model to include a singlet $s$
and 2 SU(2) doublets $\Psi_\pm$, with hypercharge $\pm 1/2$
(notice that they have the same quantum numbers as higgsinos in
the Minimal Supersymmetric Standard Model (MSSM)), with the
following Yukawa couplings:
\begin{equation}
kSss+k'S\Psi_+\Psi_-+gh^\dag \Psi_+ s+g'h\Psi_- s \label{lagrangian}
\end{equation}
There is then a reparametrization invariant CP violating phase:
\begin{equation}
\theta={\rm arg}(kk'g^*g'^*)
\end{equation}

The mass terms for this new fields can be prohibited giving proper
charges to the fields $s,\Psi_\pm$. This implies that, since at
the electroweak phase transition these new fermions get a mass of
order of the electroweak scale, and also because of the fact that
the lightest fermion is stable, we actually have a nice candidate
for dark matter. This last point is a connection between the higgs
mass, which, up to this point, we have just assured to be
exponentially smaller than the cutoff, and the weak scale, but
this connection will come out quite naturally later. So, if this
model happens to describe our universe, what we should see at LHC
should be the higgs, the two new singlets $S$ and $s$, and the two
new doublets $\Psi_{\pm}$.

Since the model is particularly minimal, it is interesting to
explore the possibility for generating the baryon number of the
universe in more detail. Before doing so, however, let us see in
more detail how the vevs of the higgs and of the singlet $S$ are
changed by the phase transition.

\subsection{Phase Transition more in detail}

Before going on, here we show more in detail how the requirement of having
a strong first order phase transition leads to have $m_S\sim m_h \sim
\Lambda$.

In unitary gauge, the equations to minimize the potential are
(from here on, we mean by $S$ also the vev of the field $S$; the
meaning will be clear by the contest):
\begin{equation}
4\lambda S^3-4\lambda_h\tilde{\lambda}S(\frac{v^2}{2} -\tilde{\lambda}S^2)+
m^2_SS+k_S\Lambda^3\Theta(T_c-T)=0\label{S_eq}
\end{equation}
\begin{equation}
2\lambda_h(\frac{v^2}{2}
-\tilde{\lambda}S^2)\frac{v}{\sqrt{2}}+2m^2_h \frac{v}{\sqrt{2}}=0
\end{equation}
where $T_c$ is the critical temperature. The first equation comes
from the derivative with respect to $S$, and we will refer to it
as $S$ equation, while for the other we will use the name $h$
equation.

We first consider the case $m^2_S>0$ and $m^2_h>0$.
Before the phase transition, we have the minimum at the symmetric vacua:
\begin{equation}
S=0,\ v=0,\ {\rm for}\ T>T_c
\end{equation}
For $T<T_c$, the minimum conditions change, and we can not solve
them analytically. We can nevertheless draw some important
conclusions. Let us consider the minimum equation for $S$, which
is the only equation which changes. Let us first consider the case
$m_S\gg\Lambda$. In this case, we can consistently neglect the
cubic terms in the $S$ equation, to get:
\begin{equation}
S\simeq-\frac{k_S\Lambda^3}{(m_S^2-2\tilde{\lambda}\lambda_h v^2)}
\end{equation}
Then, if $m_S^2\gg 2\tilde{\lambda}\lambda_h v^2$, we have:
\begin{equation}
S\simeq-\frac{k_S\Lambda^3}{m_S^2}
\end{equation}
\begin{equation}
v=0
\end{equation}
while the other solutions are still unphysical, as:
\begin{equation}
\frac{v^2}{2}\simeq-\frac{m^2_h}{\lambda_h}+\tilde{\lambda}\frac{(k^2_S\Lambda^6)}
{m_S^4}<0
\end{equation}
If $m_S^2\ll 2\tilde{\lambda}\lambda_h v^2$, then:
\begin{equation}
S\simeq-\frac{k_S\Lambda^3}{ 2\tilde{\lambda}\lambda_h v^2}
\end{equation}
For the higgs, the non null solution is:
\begin{equation}
v^2\simeq-\frac{m^2_h}{\lambda_h}
+\tilde{\lambda}\left(\frac{k^2_S\Lambda^6}
{4\tilde{\lambda}^2\lambda^2_h v^4}\right)
\end{equation}
which has some relevant effect for the electroweak phase transition 
only if $m_h\sim\Lambda$. But in that case
$v\sim\Lambda$, and from  $m_S^2\ll 2 \tilde{\lambda}\lambda_h
v^2$ we get that $m^2_S\ll \Lambda^2$, in contradiction with our
initial assumption. Then, if $\frac{m_S^2-
2\tilde{\lambda}\lambda_h v^2}{S^2}\ll 1$, in the $S$ equation, we
can consistently consider just the cubic term, to get:
\begin{equation}
S\simeq-\frac{k^{1/3}_S\Lambda}{4^{1/3}(\lambda+\tilde{\lambda}\lambda_h)^{1/3}}
\end{equation}
then, the non null solution for $v^2$ becomes:
\begin{equation}
\frac{v^2}{2}\simeq-\frac{m^2_h}{\lambda_h}
+\tilde{\lambda}\left(\frac{k^{2/3}_S\Lambda^2}
{4^{2/3}(\lambda+\tilde{\lambda}\lambda_h)(\lambda+\tilde{\lambda}\lambda_h)
}\right)
\end{equation}
which has some relevant effect only if $m^2_h\ll\Lambda^2$. However, in
this case the condition  $\frac{m_S^2- 2\tilde{\lambda}\lambda_h
v^2}{S^2}\ll 1$ would imply $\frac{m_S}{\Lambda}\ll 1$, again in
contradiction with our assumptions.

So, in order to have some effect on the higgs phase transition,
we are left with the only possibility of having $m_S \ll \Lambda$.

Restricting to this, consistently, we can neglect the linear term
in the $S$ equation, to have:
\begin{equation}
4\lambda S^3-4\lambda_h\tilde{\lambda}S\left(\frac{v^2}{2}-
\tilde{\lambda}S^2\right)=-k_S\Lambda^3
\end{equation}
\begin{equation}
2\lambda_h\left(\frac{v^2}{2}-\tilde{\lambda}S^2\right)=-2m^2_h
\end{equation}
which implies the following equation for $S$:
\begin{equation}
4 \lambda S^3+4 \tilde{\lambda}m^2_h S=-k_S\Lambda^3
\end{equation}
For $m_h\gg\Lambda$, we have:
\begin{equation}
S=-\frac{k_S\Lambda^3}{4\tilde{\lambda}m^2_h}
\end{equation}
and the non null solution for $v$ is still not physical:
\begin{equation}
\frac{v^2}{2}=-\frac{m^2_h}{\lambda_h}+\tilde{\lambda}
\left(\frac{k_S\Lambda^3}{4\tilde{\lambda}m^2_h}\right)^2<0
\end{equation}
So, finally, we have that in order to have a strong first order
phase transition triggered by the new sector, we are forced to
have $m_h\ll\Lambda$, which is what we wanted to show. In detail,
for $m_h\ll\Lambda$, we implement the following phase
transition

\begin{eqnarray}
&&S:\ 0\rightarrow-\frac{k^{1/3}_S\Lambda}{(4\lambda)^{1/3}}
\label{phase_transition}
\\ \nonumber
&&v:\ 0\rightarrow 2^{1/3}\frac{\tilde{\lambda}^{1/2}k^{1/3}_S}
{\lambda^{1/3}}\Lambda
\end{eqnarray}

Now, the final step is to show that, not only $m_h\ll \Lambda$ and
$m_S\ll\Lambda$, but that actually $m_h\sim\Lambda$ and
$m_S\sim\Lambda$. The argument for this is that, scanning in the
landscape, it will be generically much more difficult to encounter
light scalar masses, as they are fine tuned. So, in the
anthropically allowed range, our parameters shall most probably be
in the upper part of the allowed range. So, we conclude that this
model predicts $m_h\sim m_S\sim \Lambda$, as we wanted to show.

This phase transition must satisfy the requirement that the
sphalerons are ineffective ($\frac{v(T_c)}{T_c}\gtrsim 1$), after the
phase transition, which occurs at, roughly, the critical
temperature $T_c\sim\Lambda$. So:
\begin{equation}
\frac{v(T_c)}{T_c}\simeq \frac{\tilde{\lambda}^{1/2}(2k_S)^{1/3}}
{\lambda^{1/3}}\gtrsim 1 \label{phase_constraint}
\end{equation}
which can clearly be satisfied for some choices of the couplings.
In order to better understand the natural values of the ratio
$v(T_c)/T_c$ in terms of the scalar couplings, it is worth to
notice that the coupling $\tilde{\lambda}$ appears in the
lagrangian as always multiplied by $\lambda_h$. So, the coupling
which naturally tends to be equal to the other ones in the
lagrangian is $\lambda_e\equiv\lambda_h\tilde{\lambda}$. With this
redefinition, we get the constraint:
\begin{equation}
\frac{v(T_c)}{T_c}\simeq
\left(\frac{\lambda_e}{\lambda_h}\right)^{1/2}\left(\frac{2k_S}
{\lambda}\right)^{1/3}\gtrsim 1 \label{phase_constraint_2}
\end{equation}
which can clearly be satisfied with some choice of the parameters.
It is also worth to write the ratio of the vevs of $S$ and $h$ after
the phase transition:
\begin{equation}
\frac{S}{v}=\left(\frac{\lambda_h}{\lambda_e}\right)^{1/2}
\label{ratio_constraint}
\end{equation}

We can also analyze the case in which $m^2_h<0$. In this case, the
electroweak phase transition would occur before the actual strong
sector phase transition. However, since we know that in the SM the
phase transition is neither enough first order, nor enough CP
violating, we still need, in order to have baryon formation, to
require that the strong sector phase transition triggers a phase
transition in the higgs sector. It is then easy to see that all
the former discussion still applies with tiny changes, and we get
the same condition $m^2_S\sim m^2_h\sim \Lambda^2$. There is a
further check to make, though, which is due to the fact that, in
order for baryogenesis to occur, we need an unsuppressed sphaleron
rate in the exterior of the bubble. This translates in the
requirement, for $m^2_h<0$:
\begin{equation}
\frac{|m_h|}{\Lambda}\ll1
\end{equation}
which can be satisfied in some vacua.

In the next of this paper, we will always assume that these
conditions are satisfied, and the sphalerons are suppressed in the
broken phase.

Now, we are ready to treat baryogenesis in detail.

\section{Electroweak Baryogenesis}

During the electroweak phase transition, we have all the necessary
conditions to fulfill baryogenesis
\cite{Riotto:1998bt,Riotto:1999yt}. We have departure from thermal
equilibrium because of the phase transition; we have CP and C violation
because of the CP and C violating interactions; and finally we have
baryon violation because of the unsuppressed sphaleron rate. The
sphaleron rate per unit time per unit volume is in fact
unsuppressed in the unbroken phase (see
eq.(\ref{unbroken_sphaleron_rate})).

There are various different effects that contribute to the final
production of baryons. For example, CP violation can be due just
to some CP violating Yukawa coupling in the mass matrix, or it can
be mainly due to the fact that, in the presence of the wall, the
mass matrix is diagonalized with space-time dependent rotation
matrixes, which induce CP violation. Further, CP violation can be
accounted for by some time dependent effective coupling in some
interaction terms. Baryon number production as well can be treated
differently. In the contest of electroweak baryogenesis which we
are dealing with, there are mainly two ways of approaching the
problem, one in which the baryon production occurs locally where
the CP violation is taking place, so called local baryogenesis, or
one in which it occurs well in the exterior of the bubble, in the
unbroken phase, in the so called non local baryogenesis. At the
current status of the art, it appears that the non local
baryogenesis is the dominant effect, at least for not very large
 velocity of the wall $v_w$, which however is believed to be not large
because the interactions with the plasma tend to slow down the
wall considerably \cite{Dine:1992wr}, and we concentrate on this
case (see Appendix A for a brief treatment of baryogenesis in the
fast wall approximation).

In order to compute the produced baryon abundance, we follow a
semiclassical 
method developed in \cite{Huet:1995sh}.
A method based on the quantum Boltzmann 
equation and the closed time path integral formalism was developed in 
\cite{Riotto:1995hh}, making a part of the method more precise. However, 
the corrections given by applying this procedure to our case 
can be expected to be in general not very important once compared to
the uncertainties associated to our poor knowledge of certain parameters
of the electroweak phase transition, as it will become clear later, which
make the computed final baryon 
abundance reliable only to
approximately one order of magnitude \cite{Huet:1995sh}. 
Further, it is nice to note that 
our general conclusions will be quite robust under our estimated 
uncertainty in the 
computation of the baryon abundance. For these reasons, the method in  
\cite{Huet:1995sh}  represents for us the
right mixture between accuracy and simplicity which is in the
scope of our paper.

Since the method is quite contorted, let us see immediately where
the basic ingredients for baryogenesis are. Departure from thermal
equilibrium obviously occurs because of the crossing of the wall.
C violation occurs because of the $V-A$ nature of the
interactions. CP violation occurs because the CP violating phases
in the mass matrix are rotated away at two different points by two
different unitary matrix such that $U(x)^\dag U(x+dx)\neq 1$.
There could be other sources of CP violation, but, in our case,
this is the dominant one. Baryon production occurs instead well in
the exterior of the wall, in the so called non-local baryogenesis
approach, where the weak sphaleron rate is unsuppressed.

Let us anticipate the general logic. The calculation is naturally
split into two part. In the first part, we compute the sources for
CP violating charges which are due to the CP violating
interactions of the particles with the incoming wall. This
calculation will be done restricting ourself to the vicinity of
the wall, and solving a set of coupled Dirac-Majorana equations to
determine the transmission and reflection coefficients of the
particles in the thermal bath when they hit the wall, which are
different for particles and antiparticles. In the wall rest frame,
the wall is perceived as a space-time dependent mass term. This
will give rise to a CP violating current. The non null divergence
of this current will be the input of the second part of the
calculation. In this second part, we shall move to a larger scale,
and describe the plasma in a fluid approximation, where we shall
study effective diffusion equations. The key observation is that,
once the charges have diffused in the unbroken phase, thermal
equilibrium of the sphalerons will force a net baryon number. In
fact, in the presence of SU(2) not neutral charges, the
equilibrium value of the baryon number is ${\it not}$ zero:
\begin{equation}
(B+L)_{eq}=\sum_{i}c_i Q_i \label{Sph_equ}
\end{equation}
where $c_i$ are coefficients which depends on the different charges
. Once produced, the baryon number will diffuse back in the broken phase,
where, due to the suppression of the sphaleron rate, it will be practically
conserved up to the present epoch. This will end our calculation.

In the next two subsections, we proceed to the two parts of the
outlined calculation.

\subsection{CP violation sources}

We begin by computing the source for the CP violating charges, following \cite{Huet:1995sh}. We
restrict to the region very close the the wall, so that the wall
can be considered flat, and we can approximately consider the
problem as one dimensional. We consider a set of particles with
mass matrix $M(z)$ where $z$ is the coordinate transverse to the
wall, moving , in the rest frame of the wall, with energy-momentum
$E,\vec{k}$. Taken $z_0$ as the last scattering point, these
particles will propagate freely for a mean free time $\tau_i$,
when they will rescatter at the point $z_0+ \tau_i v$, where $v$
is the velocity perpendicular to the wall, $k_{tr}/E$. Now,
because of the space dependent CP violating mass matrix, these
particles will effectively scatter, and the probability of being
transmitted and reflected will be different for particles and
antiparticles. This will create a current for some charges, whose
divergence will then be the source term in the diffusion equations
we shall deal with in the next section. The effect will be
particularly large for charges which are explicitly violated by
the presence of the mass matrix, and we shall restrict to them.

We introduce $J_{\pm}$ as the average current resulting from
particles moving towards the positive and the negative $z$
direction, between $z_0$ and $z_0+\Delta$, where $\Delta= \tau v$,
and $\tau$ is the coherence time of the particles due to
interactions with the plasma \cite{Huet:1995sh}. $J_{\pm}$ are the
CP violating currents associated with each layer of thickness
$\Delta$. $J_+$ receives, for example, contributions from
particles originating from the thermal bath at $z_0$ with velocity
$v$, and propagating until $z_0+\Delta$, as well as from particles
originating at $z_0+\Delta$ with velocity $-v$, and being
reflected back at $z_0+\Delta$. The formula for $J_{\pm}$ is given
by:
\begin{equation}
J_+=\left({\rm Tr}\left(\rho_{z_0}\left(T^\dag Q T-\bar{T}^\dag Q \bar{T}\right)
\right)-{\rm Tr}\left(\rho_{z_0+\Delta}
\left(\tilde{R}^\dag Q\tilde{R}-\bar{\tilde{R}}^\dag Q \bar{\tilde{R}}
\right)\right)\right)\left(1,0,0,\tilde{v}\right)
\end{equation}
\begin{equation}
J_-=\left({\rm Tr}\left(\rho_{z_0}\left(R^\dag Q R-\bar{R}^\dag Q \bar{R}\right)
\right)-{\rm Tr}\left(\rho_{z_0+\Delta}
\left(\tilde{T}^\dag Q\tilde{T}-\bar{\tilde{T}}^\dag Q \bar{\tilde{T}}
\right)\right)\right)\left(1,0,0,-v\right)
\end{equation}
where $R(\bar{R})$ and $T(\bar{T})$ are reflection and
transmission matrices of particles (antiparticles) produced at
$z_0$ with probability $\rho_{z_0}$, evolving towards positive
$z$; while $\tilde{T}$ and $\tilde{R}$ are the correspondent
quantities for particles produced at $z_0+\Delta$ with probability
$\rho_{z_0+\Delta}$ and evolving towards negative $z$; $v$ is the
group velocity perpendicular to the wall at the point $z_0$, and
$\tilde{v}$ is the same quantity at $z_0+\Delta$; $Q$ is the
operator correspondent to the chosen charge; and the trace is
taken over all the relevant degrees of freedom and averaged over
location $z_0$ within a layer of thickness $\Delta$. When boosted
in the plasma rest frame, these currents will become the building
block to construct the CP violating source for the charges that,
diffusing into the unbroken phase, will let the production of
baryon number possible.

Now, consider a small volume of the plasma, in the plasma rest frame.
As the wall crosses it, it leaves
a current density equal to $(J_++J_-)^\mu_{plasma}$ every time interval $\tau$,
where the subscript $plasma$ refers to the quantity boosted in the plasma
frame. So, at a time $t$, the total current density accumulated will be given
by:
\begin{equation}
s^\mu=\int^t_{t-\tau_R}dt'\frac{1}{\tau}(J_+(\vec{x},t')+
J_-(\vec{x},t')))^\mu_{plasma}
\end{equation}
where $\tau_R$ is the relaxation time due to plasma interaction. From this,
the rate of change of the charge $Q$ per unit time is given by:
\begin{eqnarray}
\gamma_Q(\vec{x},t)=&&\partial_\mu s^\mu= \label{source}\\ \nonumber
&&\frac{1}{\tau}\left(J_+(\vec{x},t)+J_-(\vec{x},t)\right)^0_{plasma}
-\frac{1}{\tau}\left(J_+(\vec{x},t-\tau_R)+
J_-(\vec{x},t-\tau_R)\right)^0_{plasma}\\ \nonumber
&&-\int^t_{t-\tau_R}dz\partial_z
(J_++J_-)^z_{plasma}
\end{eqnarray}

Since in the SM CP violation is very small, and already proved to
be not enough to account for the baryon number of the universe, we
can clearly concentrate in the sector of the neutral particles of
the new theory, where
the mass matrix is given by:
\begin{displaymath}
M(z)=
\left(\begin{array}{ccc}
k S(z) & \frac{g v}{\sqrt{2}} & \frac{g' v}{\sqrt{2}} \\
\frac{g v}{\sqrt{2}} & 0 & k' S(z) \\
\frac{g' v}{\sqrt{2}} &k' S(z) & 0
\end{array}
\right) \label{mass_matrix}
\end{displaymath}
The transmission and reflection coefficients can be found by
solving the free coupled Dirac-Majorana equations for these
particles with the mass matrix given in $M(z)$. We can solve this
by a method developed in \cite{Huet:1995sh}, in a perturbative
expansion in mass insertion. As it is explained in
\cite{Huet:1995sh, Huet:1994jb}, the small parameter in the
expansion is $m\Delta$, where $m$ is the typical mass of the
particles, in the case $w\Delta<1$, or $m/w$ in the case
$w\Delta>1$. In both cases, the expansion parameter is smaller
than one. This can be understood noticing the analogy of our
system with the scattering off a diffracting medium with a step
potential of order $m$. In that case, reflection and transmission
are comparable (and this is the only case in which we produce a
net CP violating charge) only if the wave packet penetrates
coherently over a distance of order $1/m$, and has few oscillation
over that distance. Suppression of the reflection occurs both if
$m\Delta\ll1$ and if $m/w\ll1$. In the first case, this is because
only a layer of thickness $\Delta$ contributes to the coherent
reconstruction of the reflected wave, while in the second case,
because fast oscillations tend to attenuate the reconstruction of
the reflected wave. Up to sixth order in the mass insertion, we
get:
\begin{eqnarray}
T=&&1-\int^\Delta_0 dz_1\int^{z_1}_0 dz_2 M_2M^*_1 e^{2iw(z_1-z_2)}+\\
\nonumber
&& +\int^\Delta_0 dz_1\int^{z_1}_0 dz_2\int^\Delta_{z_2}dz_3\int^{z_3}_0 dz_4
M_4M^*_3M_2M^*_1 e^{2iw(z_1-z_2+z_3-z_4)}+ \\ \nonumber
&& -\int^\Delta_0 dz_1\int^{z_1}_0 dz_2\int^\Delta_{z_2}dz_3\int^{z_3}_0 dz_4
\int^\Delta_{z_4} dz_5\int^{z_5}_0 dz_6
M_6M^*_5M_4M^*_3M_2M^*_1 e^{2iw(z_1-z_2+z_3-z_4+z_5-z_6)}+\\ \nonumber
&&+ ...
\end{eqnarray}
\begin{eqnarray}
\tilde{T}=&&1-\int^\Delta_0 dz_1\int^{\Delta}_{z_1}dz_2 M^*_2M_1
e^{-2iw(z_1-z_2)}+\\ \nonumber
&& +\int^\Delta_0 dz_1\int^{\Delta}_{z_1} dz_2\int^{z_2}_0dz_3
\int^\Delta_{z_3} dz_4
M^*_4M_3M^*_2M_1 e^{-2iw(z_1-z_2+z_3-z_4)}+ \\ \nonumber
&& -\int^\Delta_0 dz_1\int^\Delta_{z_1} dz_2
\int^{z_2}_0 dz_3\int^\Delta_{z_3} dz_4
\int^{z_4}_0 dz_5\int^\Delta_{z_5} dz_6
M^*_6M_5M^*_4M_3M^*_2M_1 e^{-2iw(z_1-z_2+z_3-z_4+z_5-z_6)}+\\ \nonumber
&&+ ...
\end{eqnarray}
\begin{eqnarray}
R=&&-\int^\Delta_0 dz_1 M^*_1 e^{2iw(z_1)}+\\
\nonumber
&& +\int^\Delta_0 dz_1\int^{z_1}_0 dz_2\int^\Delta_{z_2}dz_3
M^*_3M_2M^*_1 e^{2iw(z_1-z_2+z_3)}+ \\ \nonumber
&& -\int^\Delta_0 dz_1\int^{z_1}_0 dz_2\int^\Delta_{z_2}dz_3\int^{z_3}_0 dz_4
\int^\Delta_{z_4} dz_5
M^*_5M_4M^*_3M_2M^*_1 e^{2iw(z_1-z_2+z_3-z_4+z_5)}+\\ \nonumber
&&+ ...
\end{eqnarray}
\begin{eqnarray}
\tilde{R}=&&\int^\Delta_0 dz_1 M_1
e^{-2iw(z_1)}+\\ \nonumber
&& -\int^\Delta_0 dz_1\int^{\Delta}_{z_1} dz_2\int^{z_2}_0dz_3
M_3M^*_2M_1 e^{-2iw(z_1-z_2+z_3)}+ \\ \nonumber
&& +\int^\Delta_0 dz_1\int^\Delta_{z_1} dz_2
\int^{z_2}_0 dz_3\int^\Delta_{z_3} dz_4
\int^{z_4}_0 dz_5
M_5M^*_4M_3M^*_2M_1 e^{-2iw(z_1-z_2+z_3-z_4+z_5)}+\\ \nonumber
&&+ ...
\end{eqnarray}
where $M_i=M(z_i)$.The analogous quantities for the antiparticles
are obtained replacing $M\rightarrow M^*$ in all the former formulas.

We also need to have the density matrices $\rho_{z_0}$ and
$\rho_{z_0+\Delta}$.We can choose these densities as describing thermal
equilibrium densities in eigenstates of the unbroken phase:
\begin{equation}
\rho_{z_0}={\rm Diag}\left(n_s(E,\tilde{v}),n_{\psi_+}(E,\tilde{v})
,n_{\psi_-}(E,\tilde{v})\right)
\end{equation}
where $n(E,v)$ is the Fermi-dirac distribution, boosted in the
wall frame:
\begin{equation}
n=\frac{1}{e^{\frac{\gamma_w(E-v_w k_{tr})}{T_c}}+1}
\end{equation}
and $\rho_{z_0+\Delta}=\rho_{z_0}(\tilde{v}\rightarrow-v)$. The
motivation of this is that particles  are produced in interaction
eigenstates which differ from mass eigenstates by a unitary
rotation; ignoring this, amounts at ignoring small corrections of
order $(M(z)/T_c)^2$. The choice of thermal distribution is
particularly good in the small velocity $v_w$ regime, in which we
have restricted, where the non thermal contribution is of order
$v_w$, and it induces corrections of order $v^2_w$ in the final
baryon density \cite{Huet:1995sh}.

Finally, we have to consider the charges which can play a role in generating
the baryon number. When choosing such charges, one has to consider that the
most important charges are those which are approximately conserved in the
unbroken phase, as these are the ones which can efficiently diffuse in the
unbroken phase, and induce a large generation of baryon number. Keeping this
in mind, it is easy to see that the only relevant charge in our model is
the ``higgs number''
charge, which in the same basis in which we expressed the mass matrix, for the
new CP violating sector particles,
is given by:
\begin{equation}
Q_h={\rm Diag}\left(0,0,1,-1\right)
\end{equation}

The name "higgs number" just comes from the fact that the fields
$\Psi_\pm$ have the same quantum numbers as higgsinos in the MSSM.
Now, we can substitute in the formulas for $J_+$ and $J_-$. In
order to keep some analytical expression, we decide to do a
derivative expansion
$M(z)=M(z_0)+M'(z_0)(z-z_0)+{\cal{O}}(\tau/w)^2$ and $v=\tilde{v}+
{\cal{O}}(\tau/w)^2$. This expansion is justified in the parameter
range $\tau\ll w$, which is expected to be approximately fulfilled
\cite{Huet:1994jb}.

Unlike in the case of the analogous calculation for the MSSM
\cite{Huet:1995sh} where the leading effect occurs at fourth order
in mass insertion, here the leading contribution occurs at sixth
order. We will explain later the reason of this. For the moment,
we get:
\begin{eqnarray}
J_+=&&(1,0,0,v)\times\Big(\\ \nonumber
&&+4 \int^\Delta_0 dz_1\int^{z_1}_0 dz_2\int^\Delta_{z_2}dz_3\int^{z_3}_0 dz_4
\int^\Delta_{z_4} dz_5\int^{z_5}_0 dz_6   \sin(2w(z_1-z_2+z_3-z_4+z_5-z_6))
\\ \nonumber
&&{\rm Im}\left(\rho_{z_0}Q_hM_6M^*_5M_4M^*_3M_2M^*_1\right)+\\ \nonumber
&&-4 \int^\Delta_0 dz_1\int^{z_1}_0 dz_2\int^\Delta_{0}dz_3\int^{z_3}_0 dz_4
\int^\Delta_{z_4} dz_5\int^{z_5}_0 dz_6\sin(2w(z_1-z_2-z_3+z_4-z_5+z_6))\\
\nonumber
&&{\rm Im}\left(\rho_{z_0}Q_hM_1M^*_2M_6M^*_5M_4M^*_3\right)+ \\ \nonumber
&&+4\int^\Delta_0 dz_1\int^{\Delta}_0 dz_2\int^\Delta_{z_2}dz_3
\int^{z_3}_0 dz_4
\int^\Delta_{z_4} dz_5\int^{z_5}_0 dz_6\sin(2w(z_1-z_2+z_3-z_4+z_5-z_6))\\
\nonumber
&&{\rm Im}\left(\rho_{z_0+\Delta}Q_hM^*_1M_6M^*_5M_4M^*_3M_2\right)
\Big)
\end{eqnarray}
\begin{eqnarray}
J_-=&&(1,0,0,-v)\times\Big(\\ \nonumber
&&-4 \int^\Delta_0 dz_1\int^\Delta_{z_1} dz_2\int^{z_2}_0dz_3\int^\Delta_{z_3}
dz_4
\int^{z_4}_0 dz_5\int^\Delta_{z_5} dz_6   \sin(2w(z_1-z_2+z_3-z_4+z_5-z_6))
\\ \nonumber
&&{\rm Im}\left(\rho_{z_0+\Delta}Q_hM_6M^*_5M_4M^*_3M_2M^*_1\right)+\\ \nonumber
&&+4 \int^\Delta_0 dz_1\int^\Delta_{z_1} dz_2\int^\Delta_{0}dz_3
\int^\Delta_{z_3} dz_4
\int^{z_4}_0 dz_5\int^\Delta_{z_5} dz_6\sin(2w(z_1-z_2-z_3+z_4-z_5+z_6))\\
\nonumber
&&{\rm Im}\left(\rho_{z_0+\Delta}Q_hM_1M^*_2M_6M^*_5M_4M^*_3\right)+ \\ \nonumber
&&-4\int^\Delta_0 dz_1\int^{\Delta}_0 dz_2\int^{z_2}_0dz_3
\int^\Delta_{z_3} dz_4
\int^{z_4}_0 dz_5\int^\Delta_{z_5} dz_6\sin(2w(z_1-z_2+z_3-z_4+z_5-z_6))\\
\nonumber
&&{\rm Im}\left(\rho_{z_0}Q_hM^*_1M_6M^*_5M_4M^*_3M_2\right)
\Big)
\end{eqnarray}
where the $z$ dependence in each mass matrix $M_i$ is to be understood
at linearized level.

We can substitute the results in eq.(\ref{source}), to get an expression
for the higgs charge source. In reality, if we take the relaxation
time large enough, and if we keep performing a derivative expansion,
only the first term in eq.(\ref{source}) is relevant. At first
order in the wall velocity, we get:
\begin{eqnarray}
\gamma_Q=&&\frac{\gamma_w v_w}{\tau}T_c^7{\cal{J}}\times\label{integral}\\ \nonumber
&&\Big(\sum_{i=\psi,s}\int \frac{d^3k}{2\pi^3}\frac{f(w_i\Delta)}{w^7}
v_i \frac{e^{E_i/T_c}}{(1+e^{E_i/T_c})^2}\frac{E_i}{T_c}\Big)
\end{eqnarray}
where ${\cal{J}}$ is a CP violating invariant:
\begin{equation}
{\cal{J}}=\frac{1}{4}gg'kk'(g^2-g'^2)\sin(\theta)
\end{equation}
and $f_i(w\Delta)$ is:
\begin{eqnarray}
f_\Psi(w\Delta)=&&\frac{v(z_0)^3S(z_0)}{24T_c^7}\large((S'(z_0)(9 - 24 w^2
\Delta^2 - 14 w^4 \Delta^4) v(z_0) - 9 v'(z_0) S(z_0)\\ \nonumber
&& - 2 w^2 \Delta(11 v'(z_0) \Delta(3 + w^2 \Delta^2) + \\
\nonumber &&3(5 + 2 w^2\Delta^2) v(z_0)) S(z_0) + 6 \cos(6
w\Delta) (-S'(z_0) v(z_0) + v'(z_0) S(z_0)) +\\ \nonumber && 6
\cos(4 w \Delta) (2S'(z_0)(-1 + 3 w^2 \Delta^2) v(z_0) + 2 v'(z_0)
S(z_0) + w^2 \Delta (-3 v'(z_0) \Delta + v(z_0)) S(z_0)) +\\
\nonumber && 3 \cos(2 w \Delta)(3 S'(z_0)(1 + 6 w^2 \Delta^2)
v(z_0)- 3 v'(z_0) S(z_0) + 2 w^2 \Delta(3 v'(z_0) \Delta + 4
v(z_0))S(z_0))
\\ \nonumber
&&+ 2 w(2 S'(z_0) \Delta(-3 +
14 w^2 \Delta^2) v(z_0)+(5 v'(z_0) \Delta (3 +
7 w^2 \Delta^2) +\\ \nonumber
&&3(1 + 7 w^2 \Delta^2) v(z_0)) S(z_0))
\sin(2 w \Delta)3 w (16 S'(z_0) \Delta v(z_0)
\\ \nonumber
&&+(-13 v'(z_0) \Delta +v(z_0)) S(z_0)) \sin(4 w \Delta) +
6 w \Delta) (-S'(z_0) v(z_0) + v'(z_0) S(z_0)) \sin(6 w \Delta)))
\large)
\end{eqnarray}
\begin{eqnarray}
f_s(w\Delta)=&& \frac{v(z_0)^3 S(z_0)}{48T_c^7}(S'(z_0)(15 + 12 w^2 \Delta^2 -
28 w^4 \Delta^4) v(z_0)\\ \nonumber
&& - 15 v'(z_0) S(z_0) -
        4 w^2 \Delta(v'(z_0) \Delta(48 +
                    11 w^2 \Delta^2) + \\ \nonumber
            &&  3(5 + 2 w^2 \Delta^2) v(z_0))S(z_0) +
        12 \cos(6 w \Delta)(S'(z_0) v(z_0) - v'(z_0) S(z_0)) +
        3 \cos(4 w \Delta)(S'(z_0)\\ \nonumber
                  &&(1 -
                    12 w^2 \Delta^2) v(z_0) - v'(z_0) S(z_0) +
              4 w^2 \Delta (6 v'(z_0) \Delta +
                    v(z_0)) S(z_00) + \\ \nonumber
       && 6 \cos(2 w \Delta(S'(z_0)(-5 +
                    14 w^2 \Delta^2) v(z_0) + 5 v'(z_0) S(z_0) +
              2 w^2 \Delta(5\ v'(z_0) \Delta + \\ \nonumber
                 &&   4 v(z_0)) S(z_0)) +
        12 w (S'(z_0) \Delta(-7 +
                    6 w^2 \Delta^2) v(z_0)
+ (v(z_0) +\\ \nonumber
&& \Delta(5 v'(z_0) (2 +
                                3 w^2 \Delta^2) +
                          7 w^2 \Delta v(z_0)) S(z_0))\sin(
            2 w \Delta)\\ \nonumber
&& -
        6 w(-9 S'(z_0) \Delta v(z_0)
                  +(12 v'(z_0) \Delta + v(z_0)) S(z_0))\sin(
            4 w \Delta) +
        12 w \Delta (S'(z_0) v(z_0)\\ \nonumber
&& - v'(z_0) S(z_0)) \sin(
            6 w\Delta)))
\end{eqnarray}

where $z_0$ here is a typical point in the middle of the wall. We
can estimate the coherence time as $\tau\sim (g_w T_c)^{-1}$, and so
we can take, roughly, the typical interval which is valid also in
the MSSM \cite{Huet:1995sh}: $15\leq\tau T_c\leq 35$, and in the
approximation of making the particles massless, which is still
consistent with our approximations, we can integrate numerically
the integral, and fit it linearly in $\tau$ (which is a good
approximation in the range of interest), to obtain:
\begin{eqnarray}
&&\gamma_Q(\vec{x},t)\simeq24 v_w\gamma_w \left(g g'k
k'(g^2-g'^2)\sin(\theta)\right)\times \label{sourceQ}\\ \nonumber
&&\frac{\left(1+\frac{1}{8}(\tau T_c-25)\right)}{\tau T_c}\frac{v^3S}{T_c^3}
\left(S' v-v' S\right)
\end{eqnarray}
Even though the reparametrization
invariant CP violating phase $\theta$ requires only 4 Yukawa
couplings to be present in the theory, it is somewhat puzzling
that the leading effect appears at sixth order in the couplings.
The reason is as follows. In the case $g=g'$, there is an
approximate $Z_2$ symmetry which exchanges the $\Psi_\pm$ fields.
This is only an approximate symmetry, because $\Psi_\pm$ differ by
their hypercharge. However, the hypercharge is considered a
subleading effect here, and never comes into play in our
computation, and so we do not see the breaking of this symmetry.
The $Z_2$ symmetry has, in the basis we have been using up to now,
the matrix form:
\begin{displaymath}
Z_2=
\left(\begin{array}{ccc}
1 & 0 & 0 \\
0 & 0 & 1 \\
0 & 1 & 0
\end{array}
\right) \label{mass_matrix}
\end{displaymath}
We clearly see that:
\begin{equation}
[M(z),Z_2]=0,\ [\rho(z),Z_2]=0,\ \{Q,Z_2\}=0
\end{equation}
which then implies that:
\begin{equation}
Tr\left(\rho_{z_0} TQT^\dag\right)=0
\end{equation}
and similar for the others terms in $J_+$ and $J_-$. So, in the
limit in which $g=g'$, the baryon asymmetry should vanish. Then,
since each particle needs to have an even number of mass
insertions  to be transmitted or reflected, and since the CP
violation requires at least 4 mass insertion, we finally obtain
that the coupling dependence in the CP violating source must be of
the form present in eq.(\ref{sourceQ}). From this discussion, it
is clear that in the case $g=g'$ the baryon production will be
heavily suppressed, and, from what we will see later, it will be
clear that the couplings in this case will have to be so large to
necessarily hit a Landau pole at very low energies, making the
model badly defined. We shall neglect this degeneracy of the
couplings for the next of the paper.

Now, we are ready to begin the second part of the computation.

\subsection{Diffusion Equations}

Here, we begin the second part of the calculation, still following \cite{Huet:1995sh}. We turn to
analyze the system at a larger scale, and approximate it to a
fluid. We then study the evolution of the CP violating charges due
to the presence of the sources and of the diffusion effects in the
plasma. To this purpose, we shall write a set of coupled
differential equations which include the effects of diffusion,
particle number changing reactions, and CP violating source terms,
and we shall solve them to find the various densities. We shall be
interested in the evolution of particles which carry some charges
which are approximately conserved in the unbroken phase. Near
thermal equilibrium, which is a good approximation for small
velocities, we can approximate the number density as:
\begin{equation}
n_i=k_i\mu_i T_c^2/6
\end{equation}
where $\mu_i$ is the chemical potential, and $k_i$ is a statistical factor
which is equal to 2 for each bosonic degree of freedom, and 1 for each
fermionic.

The system of differential equations simplifies a lot if we
neglect all couplings except the gauge couplings, the top quark
Yukawa coupling, and the Yukawa couplings in the new CP violating sector. From the beginning, we take
the interactions mediated from 
these last ones to be fast with respect to the typical timescale of the fluid.
We include the effect of strong sphaleron, but
neglect the one of the weak sphalerons until almost the end of the
computation. This allows us to forget about leptons. We need only
to keep track of the following populations: the top left doublet:
$Q=(t_L+b_L)$, the right top: $T=t_R$, the Higgs particle plus our
new fields $\Psi_{\pm}$:$H=h_0+\Psi_++\Psi_-$ Strong sphalerons
will be basically the only process to generate the right bottom
quarks $B=b_R$, and the quarks of the first two generations
$Q_{(1,2)L},U_R,C_R,S_R,D_R$. This implies that all these
abundance can be expressed in terms of the one of $B$:
\begin{equation}
Q_{1L}=Q_{2L}=-2U_R=-2D_R=-2S_R=-2C_R=-2B=2(Q+T)
\end{equation}
The rate of top Yukawa interaction, Higgs violating process, and
axial top number violation are indicated as
$\Gamma_y,\Gamma_h,\Gamma_m$, respectively. We take all the quarks
to have the same diffusion equation, and the same for the higgs
and the $\Psi$s. The charge abundances are then described by the
following set of differential equations:
\begin{eqnarray}
\dot{Q}=&&D_q\nabla^2Q-\Gamma_y(Q/k_Q-H/k_H-T/k_T)-\Gamma_m(Q/k_Q-T/k_T)\\
\nonumber && -6\Gamma_{ss}(2Q/k_Q-T/k_T+9(Q+T)/k_B)\\ \nonumber
\dot{T}=&&D_q\nabla^2T-\Gamma_y(-Q/k_Q+H/k_H+T/k_T) \\ \nonumber
&& -\Gamma_m(-Q/k_Q+T/k_T)+3\Gamma_{ss}(2Q/k_Q-T/k_T+9(Q+T)/k_B)\\ \nonumber
\dot{H}=&&D_h\nabla^2H-\Gamma_y(-Q/k_Q+T/k_T+H/k_H)-\Gamma_h H/k_h+\gamma_Q
\end{eqnarray}

We can restrict ourselves to the vicinity of the wall, so that we can neglect
the curvature of the surface, and assume we can
express everything in a variable
$\bar{z}=|\vec{r}+\vec{v}_w t|$. The resulting equations of motions become:
\begin{eqnarray}
v_wQ'=&&D_qQ''-\Gamma_y(Q/k_Q-H/k_H-T/k_T)-\Gamma_m(Q/k_Q-T/k_T)\label{system} \\
\nonumber && -6\Gamma_{ss}(2Q/k_Q-T/k_T+9(Q+T)/k_B)\\ \nonumber
v_wT'=&&D_q T''-\Gamma_y(-Q/k_Q+H/k_H+T/k_T) \\ \nonumber
&& -\Gamma_m(-Q/k_Q+T/k_T)+3\Gamma_{ss}(2Q/k_Q-T/k_T+9(Q+T)/k_B)\\ \nonumber
v_wH'=&&D_h H''-\Gamma_y(-Q/k_Q+T/k_T+H/k_H)-\Gamma_h H/k_h+\gamma_Q
\end{eqnarray}
We now assume that $\Gamma_y$ and $\Gamma_{ss}$ are very fast, and
we develop the result at ${\cal{O}}(1/\Gamma_y,1/\Gamma_{ss})$.
This allows to algebraically express $Q$ and $T$ in terms of $H$,
to get the following relationships:
\begin{equation}
Q=H\left(\frac{k_Q(9k_T-k_B)}{k_H(k_B+9k_Q+9k_T)}\right)
\end{equation}
\begin{equation}
T=-H\left(\frac{k_T(2k_B+9k_Q)}{k_H(k_B+9k_Q+9k_T)}\right)
\end{equation}
and, substituting back, we find the following effective
differential equation for $H$:
\begin{equation}
v_wH'=\bar{D}H''-\bar{\Gamma}H+\bar{\gamma}
\end{equation}
where the effective couplings are given by:
\begin{equation}
\bar{D}=\frac{D_q(9k_Qk_T-2k_Qk_B-2k_Bk_T)+D_hk_H(9k_Q+9k_T+k_B)}
{9k_Qk_T-2k_Qk_B-2k_Bk_T+k_H(9k_Q+9k_T+k_B)}
\end{equation}
\begin{equation}
\bar{\gamma}=\gamma_Q\left(\frac{k_H(9k_Q+9k_T+k_B)}
{9k_Qk_T-2k_Qk_B-2k_Bk_T+k_H(9k_Q+9k_T+k_B}\right)
\end{equation}
\begin{equation}
\bar{\Gamma}=(\Gamma_m+\Gamma_h)\left(\frac{(9k_Q+9k_T+k_B)}
{9k_Qk_T-2k_Qk_B-2k_Bk_T+k_H(9k_Q+9k_T+k_B)}\right)
\end{equation}

We can estimate the relaxation rates for the higgs number and the axial quark
number as \cite{Huet:1995sh}:
\begin{equation}
(\Gamma_m+\Gamma_h)\sim
\frac{4 M^2_W(T_c,z)}{21 g^2T_c}\lambda^2_t
\end{equation}
where $\lambda_t$ is the top Yukawa coupling, and $M_W$ is the $W$
boson mass, and, in order to keep analytical control, we 
approximate the source term and the relaxation term as step
functions:
\begin{eqnarray}
&&\bar{\gamma}=\tilde{\gamma},\ w>\bar{z}>0 \\ \nonumber
&&\bar{\gamma}=0,\ {\rm otherwise}
\end{eqnarray}
and
\begin{eqnarray}
&&\bar{\Gamma}=\tilde{\Gamma},\ \bar{z}>0 \\ \nonumber
&&\bar{\Gamma}=0,\ \bar{z}<0
\end{eqnarray}
For the source term $\bar{\gamma}$, we can take the avaraged value of 
expression (\ref{sourceQ}). However, due to our lack of knowledge of the 
details of the profiles of the fields during the phase transition, we can just 
approximate that expression with:
\begin{eqnarray}
\tilde{\gamma}_Q=&&24 v_w\gamma_w \left(g g'k
k'(g^2-g'^2)\sin(\theta)\right)\times \\ \nonumber
&&\frac{\left(1+\frac{1}{8}(\tau T_c-25)\right)}{\tau T_c}\frac{v^4S^2}{T_c^3\  W}
\end{eqnarray}
where we have taken $S'\sim S/W$ and $v'\sim v/W$, with $W$ the wall width, and we have assumed, 
as expected, that no
cancellation is occurring. Here $S$ and $v$ are taken to be of the order they are today.
In the approximation that $\bar{D}$ is constant, and with the boundary
conditions given by $H(\pm\infty)=0$, we have an analytical solution in the
unbroken phase \cite{Huet:1995sh}:
\begin{equation}
H={\cal{A}}e^{\bar{z}v_w/\bar{D}}
\end{equation}
where, in the limit that $\bar{D}\tilde{\Gamma}\ll v^2_w$, which is in general
applicable,
\begin{equation}
{\cal{A}}\simeq\frac{\bar{\gamma}}{\bar{\Gamma}}\left(1-e^{-2W\sqrt{\bar{\Gamma}/
\bar{D}}}\right)\label{a_prefactor}
\end{equation}
Note that diffusion of the higgs field, and so of the other charges,
in the unbroken phase, occurs for a distance of order  $z\sim\bar{D}/v_w$.

We now turn on the weak sphaleron rate, which is the responsible for
the baryon generation. The baryon density follows the following equation
of motion:
\begin{equation}
v_w\rho_B=D_q\rho''_B-\Theta(-\bar{z})3 \Gamma_{ws} n_L(\bar{z})
\end{equation}
where we have assumed that the weak sphaleron operates only in the
unbroken phase, and where $n_L$ is the total number density of left fermions.
The solution to this equation is given by, at first order in $v_w$:
\begin{equation}
\rho_B=-\frac{3\Gamma_{ws}}{v_w}\int^0_{-\infty}d\bar{z} \ n_L(\bar{z})
\end{equation}
Now, in the approximation in which all the particles in our theory are light,
we have:
\begin{equation}
k_Q=6,\  k_T=3,\  k_B=3,\ k_H=8
\end{equation}

and in the limit as $\Gamma_{ss}\rightarrow\infty$, the resulting
baryon abundance is zero \cite{Huet:1995sh}. This means that we have to go to the
next order in the strong sphaleron rate expansion. Note also that,
with this particle content, using the SM quark and higgs diffusion equation $D_q\sim 6/T_c$, $D_h\sim 110/T_c$
\cite{Joyce:1994zn}, we have  $\bar{D}\sim 96/T_c$ and
$\bar{D}\Gamma_y/v_w^2\gg2/v^2_w$, so that the assumption that the
Yukawa interaction is fast is self-consistent, since $\Gamma_y\sim
(27/2) \lambda^2_t \alpha_s T_c$ \cite{Huet:1995sh}. Finally, we take:
\begin{equation}
\Gamma_{ws}=6k\alpha^5_wT_c, \ \Gamma_{ss}=6k'\frac{8}{3}\alpha^4_s T_c
\end{equation}
where $k'$ is an order one parameter and $k\sim 20$ \cite{ew_sphal}.
To go to next order in the expansion in large $\Gamma_{ss}$, we write:
\begin{equation}
Q=H\left(\frac{k_Q(9k_T-k_B)}{k_H(k_B+9k_Q+9k_T)}\right)+\delta_Q
\end{equation}
\begin{equation}
T=-H\left(\frac{k_T(2k_B+9k_Q)}{k_H(k_B+9k_Q+9k_T)}\right)+\frac{k_T}{k_Q}
\delta_Q
\end{equation}
Substituting in ($\ref{system}$), we get:
\begin{equation}
\delta_Q=\left(\frac{D_q H''-v_wH'}{\Gamma_{ss}}\right)
\left(\frac{-k^2_Bk_Q(k_Q+2k_T)}{3k_H(9k_Q+9k_t+k_B)^2}\right)+
{\cal{O}}(1/\Gamma^2_{ss},1/\Gamma_y)
\end{equation}
Using $n_L=Q+Q_{1L}+Q_{2L}=5Q+4T=\left(\frac{5k_Q+4k_T}{k_Q}\right)\delta_Q$, we get:
\begin{equation}
n_L=7\delta_Q=-\frac{3}{112}\left(\frac{D_q H''-v_wH'}{\Gamma_{ss}}\right)
\end{equation}
Substituting the solution for the Higgs field, we finally get:
\begin{equation}
\frac{\rho_B}{s}=-\left(\frac{9{\cal{A}}\Gamma_{ws}}{112 \ s \ \Gamma_{ss}}\right)
\left(1-\frac{D_q}{\bar{D}}\right)
\end{equation}
where $s=(2\pi^2g_*/45)T_c^3\simeq55T_c^3$. Substituting our
parameters, we get:
\begin{eqnarray}
&&\frac{\rho_B}{s}\simeq5\times10^{-5}\left(\frac{k_{sp}/20}{k'_{sp}}\right)
v_w\gamma_w\frac{\left(1+\frac{1}{8}(T_c\tau-25)\right)}{(T_c\tau/25)}\\
\nonumber &&\left(g g'(g^2-g'^2)kk'\sin(\theta)\right)
\left(\frac{v}{T_c}\right)^3 \left(\frac{S}{T_c}\right)^2
\end{eqnarray}
It is worth to make a couple of small comments on the parametric dependence of this 
expression. The dependence of the coupling terms, and therefore on the vevs of $h$
and $S$, was explained in the former section. The wall width $W$ has simplified away because
the exponential in eq.(\ref{a_prefactor}) is small and can be Taylor expanded. The presence
of $\Gamma_{ss}$ in the denominator is due to the particular particle content of this model, for which
the leading term in the expansion in $1/\Gamma_{ss}$ is zero. The factor $10^{-5}$ is mainly
due to the factor of $\sim 10^{-2}$ from the entropy density, and the ratio $20\ \alpha^5_w/\alpha^4_s$, times
some other factors coming from the diffusion terms. 
For typical values of the wall velocity, we can take approximately
the SM range: $v_w\sim0.05-0.3$, while the mean free time is
$\tau\sim 20/T_c-30/T_c$. With these values, the produced baryon
number ranges in the regime:
\begin{eqnarray}
\frac{\rho_B}{s}\simeq\left(2\times 10^{-7}-3\times10^{-6}\right)
\left(g g'(g^2-g'^2)kk'\sin(\theta)\right)
\left(\frac{v}{T_c}\right)^3 \left(\frac{S}{T_c}\right)^2
\label{final_baryon}
\end{eqnarray}

This is the number which has to be equal to the baryon density at
BBN:
\begin{equation}
\left(\frac{\rho_B}{s}\right)_{{\rm BBN}}=9\times 10^{-11}
\end{equation}

For the moment, let us try do draw some preliminary 
conclusions on what this does imply on the
parameters of the model. Clearly, there are still too many
parameters which could be varied, so, as a beginning, we can set
all the Yukawa couplings in the new CP violating sector to be roughly equal to each other
$k\sim k' \sim g\sim g'$, and $v/T_c\sim1$, we then get:
\begin{equation}
\frac{\rho_B}{s}\sim\left(2\times 10^{-7}-4\times10^{-6}\right)
\left(g^6\sin(\theta)\right)\left(\frac{S}{v}\right)^2 \simeq
10^{-10}
\end{equation}
We then get the following constraint:
\begin{equation}
g^6 \left(\frac{S}{v}\right)^2\sin(\theta)\sim 10^{-4} \label{limit}
\end{equation}
It is natural, in this theory, to take the CP violating phase
to be of order 1. In this case:
\begin{equation}
g\sim g'\sim k\sim k'\sim  0.21 \left(\frac{v}{S}\right)^{1/3}
\end{equation}
From this we see that the assumption of considering the interactions mediated by these Yukawa couplings to be 
fast is justified for a large fraction
of the parameter space.
These values become lower bounds for the couplings
if we allow for the CP violating phase to be
smaller than order 1 (see eq.(\ref{limit})).


\section{Electric Dipole Moment}
The same CP violating phase which is responsible for baryogenesis,
induces an electric dipole moment through the 2-loop diagram shown
in fig.\ref{2lp}.

\begin{figure}[ht]
  \centering
    \includegraphics{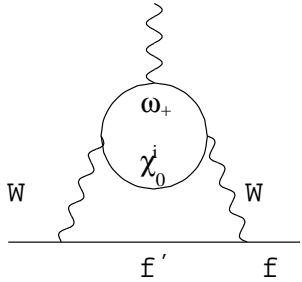}
  \caption{{\small 2 loops diagram contributing to fermion EDM, where $\chi^i_0$ are the neutral
  mass eigenstates, and $\omega_+$ is the charged one.}}
  \label{2lp}
\end{figure}

The situation here is much different than in the MSSM, where a CP
violating phase in general introduces EDM at one loop level, the constraints on which
generically force very small CP violating phases in the MSSM. It
is instead much more similar to the case of split supersymmetry
\cite{Arkani-Hamed:2004fb,Giudice:2004tc,Arkani-Hamed:2004yi},
where all the one loop diagrams contributing to the EDM are
decoupled. Here the leading diagram is at two loops level, and, as
we will soon see, it will induce electron EDM naturally just a
little beyond the present constraints, and on the edge of detection by
future experiments.

The induced EDM is (see \cite{Chang:2005ac}):

\begin{eqnarray}
\label{eq:WEDM} {d^W_f\over e} &=&  \pm \frac{\alpha^2 m_f } {8
\pi^2 s^4_W M_W^2}
 \sum_{i=1}^3  {m_{\chi_i}m_{\omega}
\over M_W^2} \hbox{ Im }(O^L_{i} O^{R*}_{i})
 {\cal G}\left( r^0_i, r^\pm, r_{f'} \right)\,,\\
 {\cal G}\left( r^0_i, r^\pm, r_{f'} \right)
 &=& \int^\infty_0 dz\int^1_0 { d\gamma \over \gamma} \int^1_0  dy\;
{y\, z\, (y +z/2 )\over (z+R)^3(z+K_{i})}\\
&=& \int^1_0 { d\gamma \over \gamma} \int^1_0 dy\, y  \left[
{(R-3K_{i})R+2(K_{i}+R)y \over 4 R(K_{i}-R)^2 }+{K_{i}(K_{i}-2y)
\over 2(K_{i}-R)^3}\ln\frac{K_{i}}{R}
 \right]\,.
\end{eqnarray}

with

 \begin{equation} R=y+(1-y)r_{f'}\,,\; K_{i}= {r^0_i \over 1-\gamma}+{r^\pm \over \gamma} \,,\;
r^\pm \equiv {m^2_\omega \over M_W^2}\,,\; r^0_i \equiv
{m_{\chi_i}^2\over M_W^2} \,,\; r_{f'} \equiv {m_{f'}^2 \over
M_W^2} \ .
\end{equation}
\begin{equation}
O^R_{i}=N_{3i}^*,\; O^L_{i}=- N_{4i} C^R \
\end{equation}

where $C^R=e^{-i\theta}$ and $N^T M_N
N=\rm{diag}\{m_{\chi_1},m_{\chi_2},m_{\chi_3}\}$ with real and
positive diagonal elements.  The plus(minus) sign on the
right-hand side of eq.(\ref{eq:WEDM}) corresponds to the fermion
$f$ with weak isospin $+(-)1/2$.  and $f'$ is its electroweak
partner. We mean by $\omega$ the charged mass eigenstate, and with
$\chi_i$, $i=1,2,3$, the neutral mass eigenstates in order of
increasing mass.

Diagonalizing the $3\times 3$ mass matrix numerically, we see that generically
the model predicts EDM very close to detection. The induced EDM
for some generic parameters are shown in fig.\ref{edm}.

\begin{figure}[ht]
  \centering
    \includegraphics[width=11.0cm,clip]{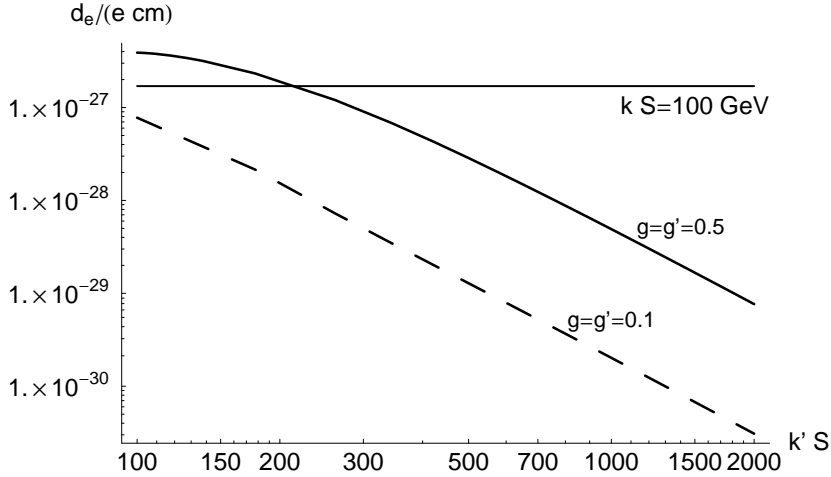}
  \caption{{\small The predicted EDM in this model. We plot the induced electron EDM
  as a function of $k'S$. The solid line represents the induced
  EDM with $g=g'=1/2$ and $k S=100$ GeV, while the dashed line represents
  $g=g'=1/10$ and $k S=100$ GeV,
  and we take maximal CP violating phase. The horizontal line represents the present electron EDM constraint
  $d_e<1.7\times 10^{-27}e \  {\rm cm}$ at $95\%$ CL \cite{edmex}.}}
  \label{edm}
\end{figure}

It is clear then that improvements of the determination of the EDM
are going to explore the most interesting region of the parameter
space. Ongoing and next generation experiments plan to improve the
EDM sensitivity by several orders of magnitude within a few years.
For example, DeMille and his Yale group~\cite{demille} will use
the molecule PbO to improve the sensitivity of the electron EDM to
$10^{-29}~e$~cm within three years, and possibly to
$10^{-31}~e$~cm within five years. Lamoreaux and his Los Alamos
group~\cite{lamo} developed a solid state technique that can
improve the sensitivity to the electron EDM by $10^4$ to reach
$10^{-31}~e$~cm. By operating at a lower temperature it is
feasible to eventually reach a sensitivity of $10^{-35}~e$~cm, an
improvement of eight orders of magnitude over the present
sensitivity. The time scale for these is uncertain, as it is tied
to funding prospects. Semertzidis and his Brookhaven
group~\cite{seme} plan to trap muons in storage rings and increase
the sensitivity of their EDM measurement by five orders of
magnitude. A new measurement has been presented by the Sussex
group~\cite{hinds}. A number of other experiments aim for an
improvement in sensitivity by one or two orders of magnitude, and
involve nuclear EDMs.

In order to understand the current and future constraints on the
model, we can study what is the induced electron EDM, once we have
satisfied the constraint from baryogenesis in eq.(\ref{final_baryon}):
\begin{equation}
g g' (g^2-g'^2)k k' \frac{v^3 S^2}{T_c^5}\sim 10^{-4}
\label{bbn_constr}
\end{equation}
The way we proceed is as follows. First, we fix $g'=g/\sqrt{2}$.
This is a good representative of the possible ratios between $g$
and $g'$, as it is quite far from the region where the approximate
symmetry in the case $g=g'$ suppresses baryogenesis, and $g'$ is
not too small to suppress baryogenesis on its own. Later on we
shall relax this condition. Having done this, we invert
eq.(\ref{bbn_constr}), to get an expression for $g$ in terms of
the other parameters of the model. Now, the couplings $g,g'$ are
expressed in terms of $T_c$, and in terms of $kS$ and $k'S$ which,
as it will be useful, in the case of small mixing, can be thought
of as respectively the singlet and the doublet mass. We shall
impose the constraint on the couplings $g,g',k,k'$ to be less than
$\sim 1$, in order for these Yukawa couplings not to hit a Landau
pole before the unification scale $\sim 10^{15}$ GeV.

We decide to restrict our analysis to the case in which the
possible ratios between the marginal couplings of the same sector
of the theory are smaller than 2 orders of magnitude. Here, by
same sector of the theory we mean either the CP violating sector,
or the sector of the scalar potential and the strong gauge group.
The justification of this relies in the fact that we wish to
explore the most natural region of the parameter space. For this
reason, we expect that there is no large hierarchy between the
marginal couplings of the same sector. We think that 2 order of
magnitudes is a threshold large enough to delimitate this natural
region. However, since marginal couplings are radiatively
relatively stable, and large hierarchies among them does not give
rise to fine tuning issues, in principle, large hierarchies among
the marginal couplings are acceptable. We think that, however,
such a hierarchy would require the addition of further structure
to the model to justify its presence, and  we decide to restrict
to the simplest realization of the model. As a consequence of
this, the most important restrictions we apply are:
$10^{-2}\lesssim k/k' \lesssim 10^2$, and $10^{-2}\lesssim
k_S/\lambda \lesssim 10^2$. In particular, using
eq.(\ref{phase_constraint_2}) and eq.(\ref{ratio_constraint}),
this implies $\frac{1}{5}\ T_c\lesssim S\lesssim 5\ T_c$, which, using
the limit $k,k'\lesssim1$, implies that the largest of $kS$ and
$k'S$ must be smaller than $5\ T_c$.

The upper bound on the CP violating sector particles tells us that
there is a small part of the parameter space which we are going to
explore, in which the particles responsible for the production of
baryons are non relativistic. In that case, we can approximately
extend the result found in eq.(\ref{bbn_constr}), with the purpose
of having order of magnitude estimates, in the following way. From
eq.(\ref{integral}), it is clear that baryon production is
Boltzmann suppressed if the particles are non relativistic. In
that case, the CP violating charge will be generated by the
scattering of these particle in the region where the induced mass
on the particles is of the order of the critical temperature, as
this is the condition of maximum CP violating interaction
compatible with not being Boltzmann suppressed. Having observed
this, it is easy to approximately extend the result of
eq.(\ref{bbn_constr}) to the non relativistic case, by taking the vevs of
the fields at the a value such that the induced mass is of the order
of the critical temperature.

In fig.3, we show the induced EDM as a function of $k S$, for
several values of the critical temperature $T_c$, for $k'S=100$, on
top, and $k'S=500$, at the bottom, with the couplings $g,g'$
chosen as explained above, in order to fulfil the baryogenesis
requirement. Notice that 500 GeV is roughly the limit that LHC
will put on SU(2) doublets. We choose the maximum CP violating
phase. The horizontal lines represent the present constraint on
EDM $d_e<1.7\times 10^{-27}e \  {\rm cm}$ at $95\%$ CL
\cite{edmex}, and the future expected one \cite{demille,lamo,seme}
of order $10^{-31}e \ {\rm cm}$. A few features are worth to be
noted. We see that, for fixed $T_c$, the EDM decreases as we
decrease $k S$, for $k S$ light enough. This is due the fact that,
reducing $k S$, we reduce
 both the EDM and the produced quantity of baryons.
However, the loss in the production of baryons is compensated with a much smaller, compared to the decrease in $k S$,
 increase in the couplings $g,g'$, so that, the baryon abundance can remain
constant, while the EDM decreases. For large $k S$, the EDM
decreases both because the mass of the particles in the loops
becomes heavier and heavier, and also because the mixing becomes
more and more suppressed. The maximum is located at the point
where $k S \sim k'S$. In that case, in fact, the mass matrix has a
diagonal piece roughly proportional to the identity, so, even
though the diagonal elements are much larger than the off diagonal
ones, mixing is much enhanced, and so is the EDM, which is
proportional to the mixing. The lower and upper limit on the value
of $k S$, as well as the minimum temperature, are dictated by the
restrictions $k' S \lesssim 5\ T_c$, $k S \lesssim 5\ T_c$, and
$k/k' \gtrsim 10^{-2}$. Now, as we decrease the critical temperature,
the resulting EDM tends to decrease. In fact, decreasing the
temperature much enhances the baryon production. This allows to
decrease the couplings $g,g'$, which explains why the induced EDM
decreases. We verify that increasing the hierarchy between $g$ and
$g'$ does not change the results a lot, as the decrease in baryons
production requires to make the other couplings large. Increasing
the value of $k' S$ up to the maximum allowed value of 1.1 TeV
(since $T^{\rm max}_c\simeq v$) does not produce any relevant change
in the result, because this raises the minimum temperature, so
that baryogenesis requires larger couplings, which forces the EDM
not to decrease relevantly. It is only once the CP violating phase
is lowered to less than $10^{-2}$ that a small experimentally
unreachable region is opened up around $k' S=500$ GeV and $k
S\lesssim 1$ GeV. The same region is opened also enlarging the
allowed hierarchy between the couplings to above $5 \times 10^2$.
However, clearly, the presence at the same time of a large hierarchy and of a very small CP
phase requires some additional structure on the model to explain
the reason for their presence.

The main conclusion we can draw from combining the analysis on
baryogenesis and EDM is that, at present, the most natural region
of the parameter space is perfectly allowed, however,
 improvements in the determination of the EDM,  are going to explore the entire viable region
of the parameter space, so that absence of signal, would result in effectively
ruling out the model, at least if no further structure is added.

\begin{figure}[ht]
  \centering
    \includegraphics[width=10.0cm,clip]{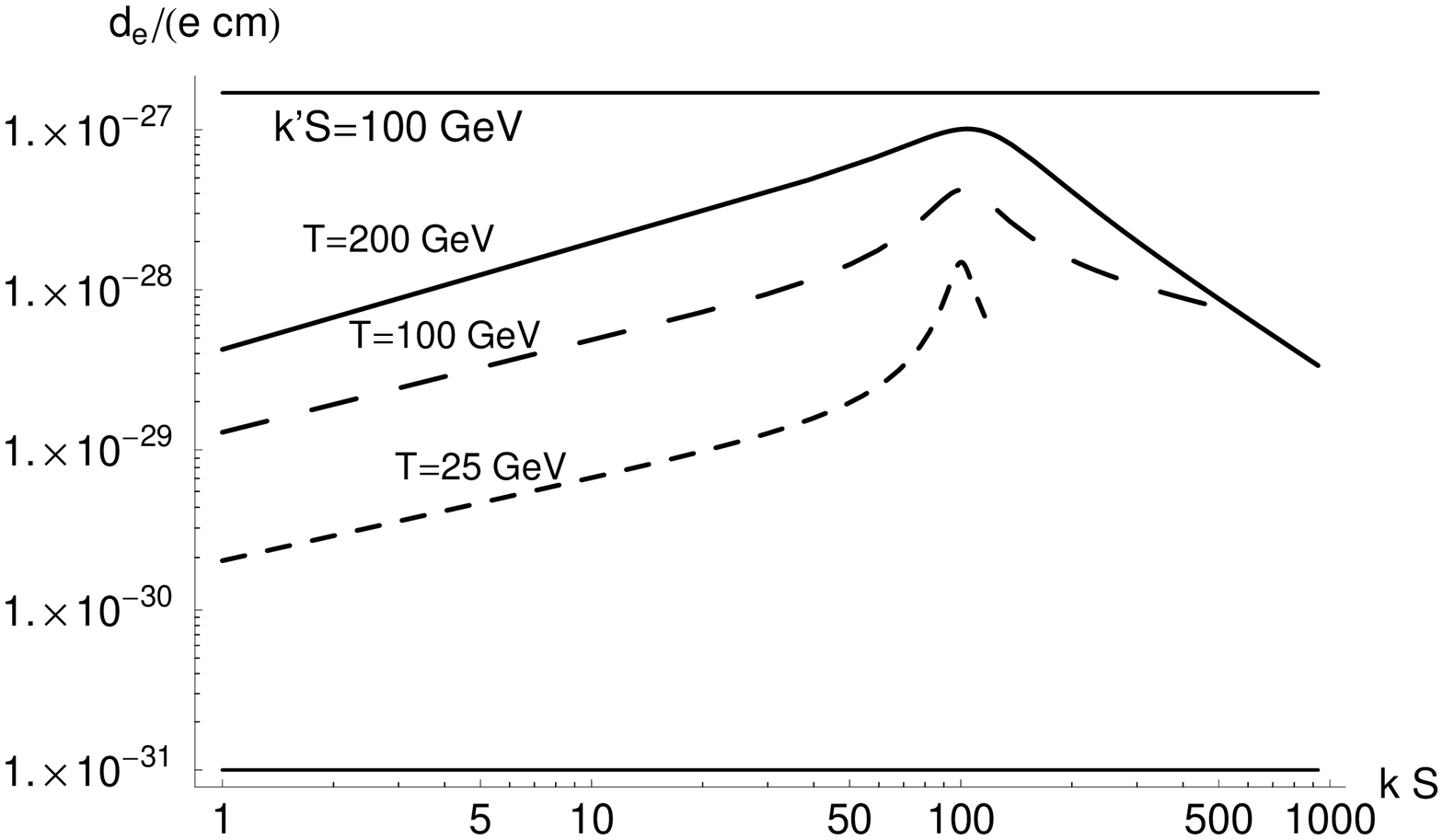}\hspace{0.3cm}
    \includegraphics[width=10.0cm,clip]{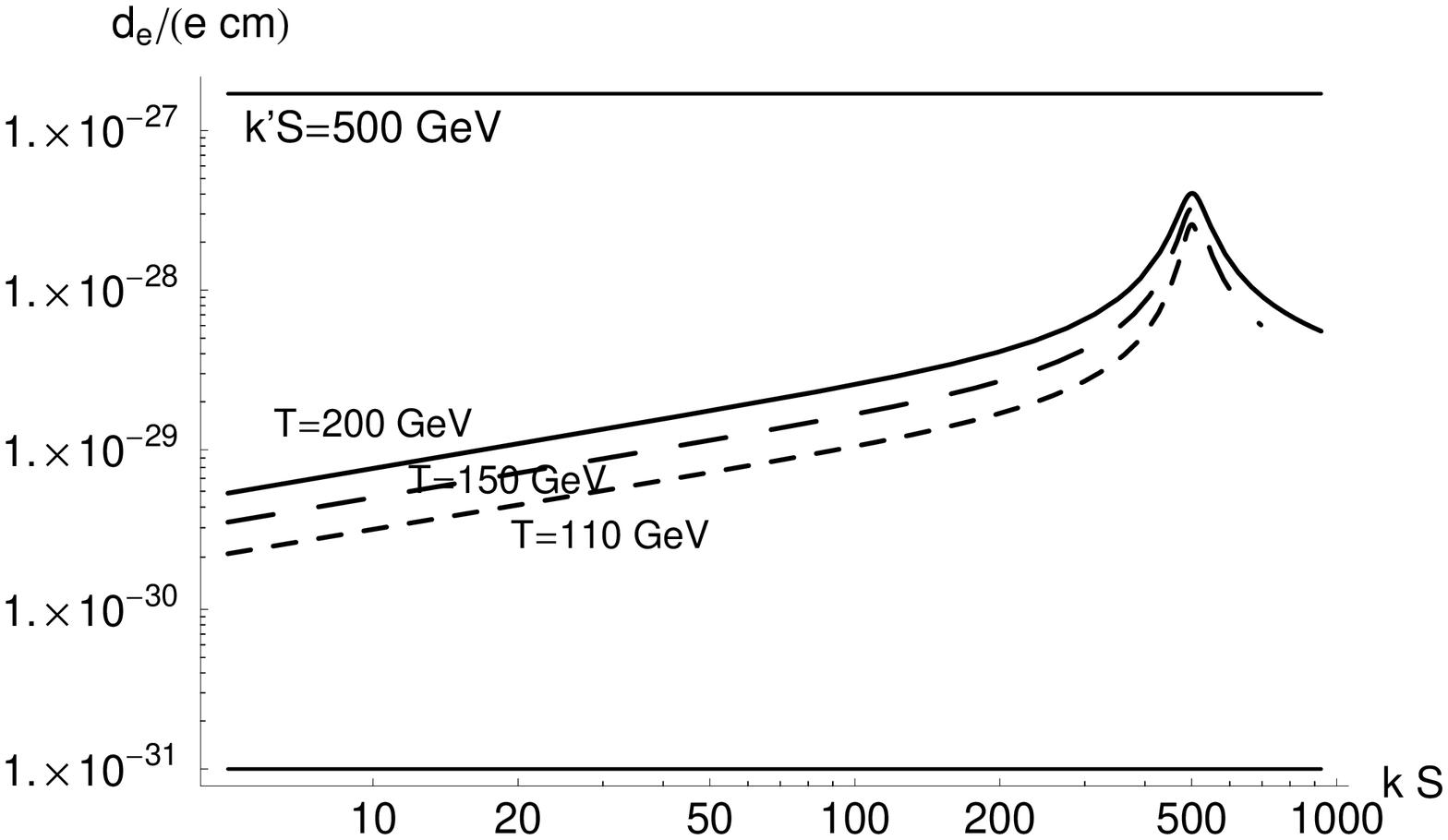}
  \caption{{\small Top, the predicted EDM given the constraint from Baryogenesis satisfied, as a function
of $k S$ for $k' S=100$ GeV,  $g'=g/\sqrt{2}$, maximum CP violating phase,
 and $T_c=200$ GeV (solid line), 100 GeV (long dashed), 25 GeV (short dashed). Bottom, the same for
$k'S=500$ GeV, and  $T_c=200$ GeV (solid line), 150 GeV (long
dashed), 110 GeV (short dashed). The horizontal lines represent
the present electron EDM constraint
  $d_e<1.7\times 10^{-27}e \  {\rm cm}$ at $95\%$ CL \cite{edmex}, and the expected improvement up to
 $d_e<10^{31}e \  {\rm cm}$ \cite{demille,lamo,seme} .}}
  \label{edm100}
\end{figure}


\section{Comments on Dark Matter and Gauge Coupling Unification}
The proposed model finds its main motivation in stabilizing the
weak scale through the requirement of attaining baryons in our
universe. However, further than this, it is clear that the model
provides two other interesting phenomenological aspects: gauge
coupling unification and dark matter. Here, we just briefly
introduce these aspects, and we postpone a more detailed
discussion to future work.

Thanks to the two doublets we have inserted in our model, which
have the same quantum numbers as higgsinos in the MSSM, gauge
coupling unification works much better than in the standard model.
As it was shown in \cite{Arkani-Hamed:2005yv}, gauge coupling
unification with the standard model plus higgsinos works roughly as well
as the MSSM at two loops level, with the only possible problem being
the fact that the unification scale is a bit low at around $\sim
10^{15}$ GeV. The problem from proton decay can however be avoided
with some particular model at the GUT scale
\cite{Arkani-Hamed:2005yv,preparation}. We expect that 2-loop
gauge coupling unification works quite well also in this model,
with only small corrections coming from the presence of the
singlet scalar $S$.

Concerning the Dark Matter relic abundance, the lightest of the
newly introduced particles is stable, and so it provides a natural
candidate for Dark Matter. Estimating the relic abundance is quite
complex, as it depends on the composition of the particle, and on
its annihilation and coannihilations rate. However, if our newly
introduced Yukawa couplings are not very close to their upper
limit (of order one), and if the lightest particle is mostly
composed of the two doublets, then its relic abundance is very
similar to the one of pure higgsino dark matter in split
supersymmetry
\cite{Arkani-Hamed:2004fb,Giudice:2004tc,Pierce:2004mk,Arkani-Hamed:2004yi,Senatore:2004zf,Masiero:2004ft},
requiring the doublets to be around the TeV scale. In the case of
singlet dark matter, we expect the singlet to give the right
abundance for a much lighter mass, as it is naturally much less
interacting. However, estimates become much more difficult, and we
postpone the precise determination of the relic abundance to
future work.


\section{Conclusions}

In this paper, we have addressed the solution to the electroweak
hierarchy problem in the context of the landscape, following
a recent model proposed in \cite{Arkani-Hamed:2005yv}.

We have shown that it is possible to connect the electroweak scale
to a hierarchically small scale at which a gauge group becomes
strong by dimensional transmutation. The assumption is that we are
in a "friendly neighborhood" of the landscape in which only the
relevant parameters of the low energy theory are effectively
scanned. We then realize the model in such a way that there is a
fragile, though necessary, feature of the universe which needs to
be realized in our universe in order to sustain any sort of life.
As a natural continuation of Weinberg's "structure principle", the
fragile feature is the presence of baryons in the universe, which
are a necessary ingredient for the formation of clumped
structures. We assume  that, in the friendly neighborhood of the
landscape in which we should be, baryogenesis is possible only
through the mechanism of electroweak baryogenesis. Then, in order
to produce a first order phase transition strong enough so that
sphalerons do not wash out the produced baryon density, we develop
a new mechanism to implement the electroweak phase transition. We
introduce a new gauge sector which becomes strong at an
exponentially small scale through dimensional transmutation. We
couple this new sector to a singlet $S$ which is then coupled to
the higgs field. The electroweak phase transition occurs as the
new gauge sector becomes strong, and produces a chiral
condensation. This triggers a phase transition for the singlet
$S$, which then triggers the phase transition for the higgs
fields. In order to preserve baryon number, we need the phase
transition to be strong enough, and this is true only if the higgs
mass is comparable to the QCD scale of the strong sector. This
solves the hierarchy problem. In order to provide the necessary CP
violation, we introduce 2 SU(2) doublets $\Psi_\pm$ with
hypercharge $\pm 1/2$, and a gauge singlet $s$.

When we require the model to describe our world, the model
leads to falsifiable predictions.

The main result of the paper is the computation of the produced baryon number,
for which we obtain:
\begin{eqnarray}
\frac{\rho_B}{s}\simeq\left(2\times 10^{-7}-3\times10^{-6}\right)
\left(g g'(g^2-g'^2)kk'\sin(\theta)\right)
\left(\frac{v}{T_c}\right)^3 \left(\frac{S}{T_c}\right)^2
\end{eqnarray}
The requirement that this baryon abundance should cope with the
observed one leads to a lower bound on a combination of the
product of the CP violating phase, the new couplings, the $S$
vev, and the critical temperature $T_c$.

We infer that Gauge Coupling Unification, and the right amount of
Dark Matter relic abundance are easily achieved in this model.

We study in detail the induced electron Electric Dipole Moment
(EDM), and we find that, at present, the most natural region of
the parameter space of the theory is allowed. However, soon in the
future improvement in the EDM experiments
 will be sensitive the entire viable region of the
parameter space of the model, so that absence of a signal
would result in practically ruling out the model.

\section*{Acknowledgments}
I would like to thank  Nima Arkani-Hamed, who inspired me the
problem, and without whose constant help I would have not been
able to perform this calculation. I would like to thank also Paolo
Creminelli, Alberto Nicolis, Toby Wiseman, and Rakhi Mahbubani for
interesting conversations.

\appendix

\section{Baryogenesis in the Large Velocity Approximation}
The method we have used in the main part of this work is based on some approximations
that fail in limit of very fast wall speed: $v_w\sim1$. Even
though this regime seems to be disfavored by actual computations
of the wall speed\cite{Dine:1992wr}, it is worth to try to
estimate the result even in this case. In this appendix
 we are going to do an approximate computation
in the large wall velocity approximation. In this new regime,
calculations become much more complicated, as the assumption of
local thermodynamical equilibrium begins to fail, and the
baryon number tends to be produced in the region of the wall, in
the so called local baryogenesis scenario.

We follow the
treatment of \cite{Lue:1996pr}. In this approach, the phase
transition is treated from an effective field theory point of view. A
good way of looking at the high temperature phase of the unbroken phase
 is imagining it as discretized in a lot of cells, whose
side is given by the typical size of the weak sphaleron barrier
crossing configuration of the gauge field: $\zeta\sim(\alpha_W
T_c)^{-1}$. Concentrating on each cell, we have that the thermal
energy in a cell is in general much larger than the energy
necessary to create a gauge field oscillation capable of crossing
the barrier, and this means that most of the energy is in
oscillations with smaller wavelength than $\zeta$. So, these
configurations cross the barrier at energy far above the one of
the sphaleron configuration, and so their rate has nothing to do
with the sphaleron rate, and this explain why their rate per unit
volume is of order $\zeta^4$ (In the case $m^2_h<0$, electroweak
symmetry breaking has already occurred outside the bubble, but
this discussion still roughly applies). We can parameterize the
configuration in one cell with one variable $\tau$, which depends
only on time. Obviously, this is a very rough approximation, but
this is at least a beginning. We try to describe the dynamics of the
configuration near crossing the barrier, at a maximum of the
energy, that we fixe to be at $\tau=0$. Near this point, we can
write the following Lagrangian, which is very similar to the one
in \cite{Lue:1996pr}:
\begin{equation}
L(\tau,\dot{\tau})=\frac{c_1}{2\zeta}\dot{\tau}^2+\frac{c_2}{2\zeta^3}\tau^2+
\frac{c_3}{\zeta}b \phi\dot{\tau}\label{effective_action}
\end{equation}
where $c_i$ are dimensionless parameters depending on the
different possible barrier crossing trajectories, while $b$ will
be determined shortly. Since the point $\tau=0$ is a maximum of
the energy, no odd powers of $\tau$ can appear. The appearance of
the odd power in $\dot{\tau}$ can be understood because,
 at one loop level, the CP violating mass matrix of the particles in the CP
violating sector induces
an operator which contains the term ${\rm Tr}F_{\mu\nu}\tilde{F}^{\mu\nu}$. This can be seen
in the following way, in an argument similar
to the one shown in \cite{Arkani-Hamed:2004yi}.
We can imagine to do a chiral rotation on these field of amplitude equal to the CP violating phase. Because
of the anomaly, this will induce the operator:
\begin{equation}
{\cal{O}}=\frac{g^2_2}{16\pi^2}\phi
{\rm Tr}F_{\mu\nu}\tilde{F}^{\mu\nu}
\end{equation}
where $g_2$ is the SU(2) weak coupling, and $F_{\mu\nu}$ is the SU(2)
field strength, and where
\begin{equation}
\phi={\rm{Arg(Det(}}M))={\rm Arcsin}
\left(\frac{k k' S^2{\rm sin}(\theta)}{k k' S^2{\rm sin}(\theta)^2+(g g' v^2+k k' S^2 {\rm cos}(\theta))^2}\right)
\end{equation}
Promoting the vevs of $h$ and $S$ to the actual fields, we find the operator we were looking for. 
Now, the term
${\rm Tr}F_{\mu\nu}\tilde{F}^{\mu\nu}$ contains a term proportional to the
time derivative of the Chern Simons number, and so the operator
${\cal{O}}$ must contribute to the effective action
($\ref{effective_action}$) with a term proportional to
$b \phi\dot{\tau}$, with $b=\frac{g^2_2}{16 \pi^2}$, explaining the 
reason for the presence of the term in $\dot{\tau}$

The equation of motion for $\tau$ is:
\begin{equation}
\ddot{\tau}=\frac{c_2}{c_1\zeta^2}\tau-\frac{c_3}{c_1}
\frac{g^2_2}{16\pi^2}
\dot{\phi}
\end{equation}
If the wall is fast enough,we can solve it in the impulse approximation,
to get:
\begin{equation}
\Delta\dot{\tau}=-\frac{c_3}{c_1}\frac{g^2_2}{16\pi^2}
\Delta\phi
\end{equation}
This kick to $\Delta\dot{\tau}$ makes the distribution of velocities of
barrier crossing configurations asymmetric, leading to a
production of baryons with respect to antibaryons.

Now, this kick will be very inefficient in changing the
distribution of baryons, unless the kick is larger than the typical speed
$\dot{\tau}_0$ a generic configuration would have if it crossed
the barrier in the absence of the wall. So, we require:
$\Delta\dot{\tau}>\dot{\tau}_0$. The fraction of configurations
which satisfy this requirement is proportional to
$\Delta\dot{\tau}$, but is very difficult to estimate. Following
\cite{Lue:1996pr}, we just say
that it is equal to $f \Delta\dot{\tau}$, where $f$ is our
``ignorance'' coefficient. So, we finally get:
\begin{equation}
n_B\sim f\Delta\dot{\tau}\zeta^{-3}
\end{equation}
where we have reabsorbed the constants $c_i$ into $f$. We finally
get:
\begin{equation}
\frac{\rho_B}{s}\sim f
\frac{\alpha^3_W}{45}\frac{g^2_2}{16\pi^2}\Delta\phi\label{rajo}
\end{equation}
There are a lot of heavy approximations which suggest that this
estimate is very rough \cite{Lue:1996pr}: from the value of the
coefficient $c_i$, to the approximation of restricting to one
degree of freedom, to the impulse approximation in solving the
differential equation, and to the estimate of the fraction of
configurations influenced by the kick. All these suggests that we
should take eq.({\ref{rajo}) with $f\sim 1$, as at most an upper
limit on the baryon production, as the authors of
\cite{Lue:1996pr} suggest.

\end{document}